\documentstyle[12pt,amsmath,amssymb,hyperref,color, graphics, appendix]{article}

\makeatletter \@addtoreset{equation}{section}

\makeatother

\makeatletter
\newcommand\appendix@section[1]{%
  \refstepcounter{section}%
  \orig@section*{Appendix \@Alph\c@section: #1}%
  \addcontentsline{toc}{section}{Appendix \@Alph\c@section: #1}%
}
\let\orig@section\section
\g@addto@macro\appendix{\let\section\appendix@section}
\makeatother

\def\p{\partial}
\def\be{\begin{eqnarray}}
\def\ee{\end{eqnarray}}
\def\nn{\nonumber}
\def\l{\langle}
\def\r{\rangle}
\def\p{\partial}
\DeclareMathOperator*{\ress}{res}
\newcommand{\beq}{\begin{equation}}
\newcommand{\eeq}{\end{equation}}

\voffset= - 1.3in
\hoffset= - 0.8in         

\title{{\bf Minimal Liouville Gravity correlation numbers from Douglas string equation.}}
\vspace{.2cm}

\begin{document}

\author{Alexander Belavin$^{1,2,3}$, Boris Dubrovin$^{4,5,6}$, Baur Mukhametzhanov$^{1,7}$\footnote{E-mail:  baur@itp.ac.ru} 
\vspace*{10pt}\\[\medskipamount]
$^1$~\parbox[t]{0.88\textwidth}{\normalsize\it\raggedright
L.D.Landau Institute for Theoretical Physics,
142432 Chernogolovka, Russia}
\vspace*{10pt}\\[\medskipamount]
$^2$~\parbox[t]{0.88\textwidth}{\normalsize\it\raggedright
Moscow Institute of Physics and Technology, 141700 Dolgoprudny, Russia}
\vspace*{10pt}\\[\medskipamount]
$^3$~\parbox[t]{0.88\textwidth}{\normalsize\it\raggedright
Institute for Information Transmission Problems, 127994, Moscow, Russia}
\vspace*{10pt}\\[\medskipamount]
$^4$~\parbox[t]{0.88\textwidth}{\normalsize\it\raggedright
International School of Advanced Studies (SISSA), 
34136 Trieste, Italy}
\vspace*{10pt}\\[\medskipamount]
$^5$~\parbox[t]{0.88\textwidth}{\normalsize\it\raggedright
N.N.Bogolyubov Laboratory for Geometrical Methods in Mathematical Physics, Moscow State University ``M.V.Lomonosov", 119899 Moscow, Russia}
\vspace*{10pt}\\[\medskipamount]
$^6$~\parbox[t]{0.88\textwidth}{\normalsize\it\raggedright
V.A.Steklov Mathematical Institute, 119991 Moscow, Russia}
\vspace*{10pt}\\[\medskipamount]
$^7$~\parbox[t]{0.88\textwidth}{\normalsize\it\raggedright
Department of Physics, Harvard University, 02138 Cambridge, USA}
}

\baselineskip 0.6cm

\date{}
\maketitle
 \vspace{-0.9cm}

\vspace{1cm}

\begin{abstract}
We continue the study of $(q,p)$ Minimal Liouville Gravity with the help of Douglas string equation. We generalize the results of \cite{Moore:1991ir}, \cite{Belavin:2008kv}, where Lee-Yang series $(2,2s+1)$ was studied, to $(3,3s+p_0)$ Minimal Liouville Gravity, where $p_0=1,2$. We demonstrate that there exist such coordinates $\tau_{m,n}$ on the space of the perturbed Minimal Liouville Gravity theories, in which the partition function of the theory is determined by the Douglas string equation. The coordinates $\tau_{m,n}$ are related in a non-linear fashion to the natural coupling constants $\lambda_{m,n}$ of the perturbations of Minimal Lioville Gravity by the physical operators $O_{m,n}$. We find this relation from the requirement that the correlation numbers in Minimal Liouville Gravity must satisfy the conformal and fusion selection rules. After fixing this relation we compute three- and four-point correlation numbers when they are not zero. The results are in agreement with the direct calculations in Minimal Liouville Gravity available in the literature \cite{Goulian:1990qr}, \cite{Zamolodchikov:2005sj}, \cite{Belavin:2006ex}. 
\end{abstract}


\newpage

\baselineskip 0.6cm

\maketitle

\tableofcontents

\newpage

\section{Introduction}
${}$

Since the invention of the Minimal Liouville Gravity \cite{Polyakov:1981rd}, \cite{Knizhnik:1988ak}, one of its important problems was to explicitly find arbitrary $n$-point correlation functions of physical operators in this theory. The Minimal Liouville Gravity represents probably the simplest example of two-dimensional quantum gravity, when the matter sector is taken to be a $(q,p)$ Minimal Model of CFT \cite{Belavin:1984vu}. However even in this theory it turned out to be a highly non-trivial problem to calculate $n$-point correlation functions. 
The main difficulty is to calculate the integrals over the moduli space of Riemann surfaces with $n$ punctures. Some progress in this direction was achieved in \cite{Belavin:2006ex}, where, using higher Liouville equations of motion \cite{Zamolodchikov:2003yb}, a new technique for this sort of integrals was developed by Alexei Zamolodchikov and one of the authors. In this way the 4-point functions on the sphere in Minimal Gravity were calculated there \cite{Belavin:2006ex}.

The most natural physical objects to calculate in Minimal Liouville Gravity are the $n$-point correlation functions

\be 
Z_{m_1n_1\dots m_Nn_N} =  \l O_{m_1n_1} \dots O_{m_Nn_N} \r, \qquad O_{m,n} = \int_M {\cal O}_{m,n}
\ee 
where ${\cal O}_{m,n}$ -- are some local physical observables which are to be introduced in the section 2. We also restrict ourselves to the case when the manifold $M$ has the topology of the sphere. Often the quantities $Z_{m_1n_1\dots m_Nn_N}$ are referred to as correlation numbers. It is convenient to consider the generating function

\begin{align} \label{introduction gen function}
Z_L(\lambda) &= \l \exp \sum_{m,n} \lambda_{m,n} O_{m,n} \r, \\
Z_{m_1n_1\dots m_Nn_N} &= {\p \over \p \lambda_{m_1n_1} } \dots {\p \over \p \lambda_{m_Nn_N} } \Bigg|_{\lambda =0} Z_L(\lambda) 
\label{intro corr numb}
\end{align}
The generating function (\ref{introduction gen function}) can be understood as the partition function of the perturbed theory. For that reason we call it partition function. We can think of the coupling constants $\lambda_{m,n}$ as of the coordinates on the space of the perturbed Minimal Liouville Gravity theories.

On the other hand since the late 1980s another, discrete, 
approach \cite{Kazakov:1985ea}, \cite{Kazakov:1986hu}, \cite{Kazakov:1989bc}, \cite{Staudacher:1989fy}, \cite{Brezin:1990rb}, \cite{Douglas:1989ve}, \cite{Gross:1989vs}
to 2-dimensional gravity (as opposed to the continuous approach of Minimal Liouville Gravity) has been developing.
\footnote{There exists the third approach, called topological gravity, but we do not aim to discuss it in this paper.}
In this approach a fluctuating 2-dimensional surface is approximated by an ensemble of graphs. The continuous geometry is restored in the scaling limit, when large size graphs dominate. This approach is realized through Matrix Models (find a survey of these ideas in \cite{Ginsparg:1993is}, \cite{DiFrancesco:1993nw}).    

Since both approaches are based on the same idea of 2-dimensional fluctuating geometry, they were expected to give the same results for physical quantities. This was checked by calculations in some particular models \cite{Knizhnik:1988ak}, \cite{Goulian:1990qr}, \cite{DiFrancesco1992}.


In 1989 Douglas in his seminal paper \cite{Douglas:1989dd} had shown that the partition function in the discrete approach satisfies a differential equation and the so-called "string equation", which is usually referred to as Douglas string equation. The times $\tau_{m,n}$ in the generalized KdV hierarchy play a role of the perturbation parameters $\lambda_{m,n}$ in Minimal Gravity. Indeed, it was shown in \cite{Douglas:1989dd} that the scaling dimensions of the times $\tau_{m,n}$ coincide with those of the coupling constants $\lambda_{m,n}$ in Minimal Gravity. However a na\"{\i}ve identification of the times $\tau_{m,n}$ and coupling constants $\lambda_{m,n}$ leads to inconsistencies. Namely, in Minimal Gravity one has conformal and fusion selection rules. For instance, the one-point correlation numbers of all operators, except the unity operator, must be zero, the two-point correlation numbers must be diagonal and there are similar restrictions to higher point correlation numbers. These conformal and fusion selection rules are not satisfied in the Douglas approach if we na\"{\i}vely identify $\tau_{m,n}$ and $\lambda_{m,n}$. This problem was first pointed out by Moore, Seiberg and Staudacher in \cite{Moore:1991ir}. There they also proposed how this problem can be solved. The idea of \cite{Moore:1991ir} was that due to possible contact terms in the correlators, the times $\tau_{m,n}$ in the Douglas approach and coupling constants $\lambda_{m,n}$ in Minimal Liouville Gravity are related in a non-linear fashion like

\be \label{intro resonance}
\tau_{m,n} = \lambda_{m,n} + \sum_{m_1n_1m_2n_2} C_{m,n}^{m_1n_1m_2n_2} \lambda_{m_1n_1}\lambda_{m_2n_2} + \dots
\ee
Appropriate choice of this substitution allowed them to explicitly establish the correspondence up to two-point correlation numbers in $(2,2s+1)$ Minimal Gravity. 

After the 3- and 4-point correlation numbers in Minimal Gravity had been calculated in \cite{Goulian:1990qr}, \cite{Zamolodchikov:2005sj}, \cite{Belavin:2006ex}, it became possible to make more explicit checks against Douglas equation approach. These kind of checks were performed in \cite{Belavin:2008kv} for $(2,2s+1)$ Minimal Gravity, where, using the ideas of \cite{Moore:1991ir}, the full correspondence between Douglas equation approach and Minimal Gravity was proposed for arbitrary $n$-point correlation numbers.

The aim of this paper is to generalize the results of \cite{Moore:1991ir}, \cite{Belavin:2008kv} to the case of $(3,3s+p_0)$ Minimal Gravity.
\footnote{In $(q,p)$ Minimal CFT the numbers $q$ and $p$ must be mutually prime \cite{Belavin:1984vu}.}
Our analysis is based on the following assumptions:
\vspace{0.3cm}

$\bullet$ There exist special coordinates $\tau_{m,n}$ in the space of perturbed Minimal Liouville Gravities such that the partition function of Minimal Liouville Gravity satisfies, as in matrix models, a partial differential equation and the Douglas string equation.

$\bullet$ Besides, one can show that the solution of the Douglas string equation satisfies the generalized KdV hierarchy equations \cite{Krichever:1992sw}, \cite{Dubrovin:1992dz}.

$\bullet$ The times $\tau_{m,n}$ are related to the natural coordinates $\lambda_{m,n}$ (\ref{introduction gen function}) on the space of the perturbed Minimal Liouville Gravities by a non-linear transformation like in (\ref{intro resonance}). We establish the form of this transformation $\tau(\lambda)$ by the requirement that the correlation numbers (\ref{intro corr numb}), which are the coefficients of the expansion of the partition function in the coordinates $\lambda_{m,n}$, satisfy the conformal and fusion selection rules.

$\bullet$ The differential equation for the partition function and the Douglas string equation define the partition function as logarithm of Sato's tau-function \cite{Segal1985} of the dispersionless generalized KdV hierarchy \cite{Krichever:1992sw}, \cite{Dubrovin:1992dz} with initial conditions defined by the Douglas string equation.

\vspace{0.3cm}

To simplify our analysis we need an explicit expression for the partition function. We find such an expression using the relation of the Douglas string equation and dispersionless KdV hierarchy with the Frobenius manifold structure \cite{Dijkgraaf:1990dj}, \cite{Dubrovin:1992dz}. This relation leads to a convenient integral representation of the partition function (\ref{free energy integral}) for all $(q,p)$. Using this expression we analyse the case $(3,3s+p_0)$ where $p_0=1,2$. As in \cite{Moore:1991ir} and \cite{Belavin:2008kv} the relation between Minimal Gravity and Douglas equation approach is found to be non-trivial because of the contact terms in the correlation functions. Namely, the scaling dimensions of the coupling constants $\lambda_{m,n}$ in (\ref{introduction gen function}) coincide with those of the times $\tau_{m,n}$ in the Douglas equation approach. However, the relation between them is found to be non-linear. This relation is established by the requirement that there exists a polynomial $\tau_{m,n}(\{\lambda_{m,n}\})$ such that after the substitution $\tau_{m,n} \to \tau_{m,n}(\{\lambda_{m,n}\})$ the partition function in the Douglas equation approach coincides with the partition function (\ref{introduction gen function}) in Minimal Gravity and the obtained correlation numbers satisfy the conformal and fusion selection rules. In this way the relation $\tau(\lambda)$ is implicitly established in terms of the Jacobi polynomials $P_n^{(a,b)}$ up to the third powers of $\lambda$ in (\ref{intro resonance}). Also, after establishing the relation $\tau(\lambda)$, we evaluate the correlation numbers when they are non-zero. The results are in agreement with the direct evaluations of three-point \cite{Goulian:1990qr}, \cite{Zamolodchikov:2005sj} and four-point \cite{Belavin:2006ex} correlation numbers in Liouville Minimal Gravity. 

In spite of the agreement in the non-zero three- and four-point correlation numbers, using the substitution $\tau(\lambda)$ in the partition function in the Douglas equation approach we managed to satisfy only a part of the conformal and fusion selection rules, while the other part remains unsatisfied. We will discuss this issue in the final parts of the paper.

In section 2 we review the conformal and fusion selection rules and the results for three- and four-point correlation numbers in Minimal Gravity. Also we discuss in this section the subtlety of contact terms which makes the relation $\tau(\lambda)$ non-trivial. Then in section 3 we discuss the Douglas approach to Minimal Gravity. In section 4 we warm up by rederiving the results of the paper \cite{Belavin:2008kv} for $(2,2s+1)$ Minimal Gravity in a slightly different way. And finally, we make generalization to the case $(3,3s+p_0)$ in section 5. The concluding remarks and discussion of the results can be found in section 6. To make the text self-consistent we give a review of ideas on Frobenius manifolds in Appendix A.


\section{Minimal Liouville Gravity}
The Liouville Gravity consists of Liouville theory of a scalar $\phi$ and some conformal matter sector. One of the simplest cases is when the matter sector is taken to be a $(q,p)$ Minimal Model of CFT \cite{Belavin:1984vu}. This theory is called Minimal Liouville Gravity.


\subsection{Minimal Models}

The Minimal Models of CFT ${\cal M}_{q,p}$, labelled by two co-prime integers $(q,p)$, have a finite number of primary fields, which are enumerated by the Kac table: $\Phi_{m,n}$, where $m=1,\dots, q-1$ and $n=1,\dots, p-1$. Due to the relation 

\be 
\Phi_{m,n}= \Phi_{q-m, p-n}
\ee
only a half of the fields $\Phi_{m,n}$ are independent.

In this paper we consider the Minimal Models ${\cal M}_{2,2s+1}$ and ${\cal M}_{3,3s+p_0}$, where $p_0 = 1,2$. In these models it suffices to consider the fields from the first raw $\Phi_{1,k}$. Let us introduce a notation 

\be 
\Phi_k = \Phi_{1,k+1}.
\ee
In the model $(2,2s+1)$ the independent fields are $\Phi_k$, $0 \leq k \leq s-1$. In the model $(3,3s+p_0)$ the independent fields are $\Phi_k$, $0 \leq k \leq p-2$.  

The operator product expansion (OPE) for these fields is

\be \label{MM OPE}
[\Phi_{k_1}] [\Phi_{k_2}] = \sum_{k=|k_1-k_2|:2}^{I(k_1,k_2)} [\Phi_k]
\ee
where, as usually, $[\Phi_k]$ denotes the contribution of the irreducible Virasoro representation with the highest state $\Phi_k$. Here the symbol $k=|k_1-k_2|:2$ denotes summation that goes from $k=|k_1-k_2|$ till $I(k_1, k_2)$ with the step 2 where
\be
I(k_1,k_2) = \min (k_1+k_2, 2p-k_1-k_2-4). 
\ee
The small conformal group of fractional-linear transformations together with OPE (\ref{MM OPE}) puts strong constraints on correlation functions. Particularly, many of them are zero. For instance the small conformal group constraints one- and two- point correlation functions

\begin{align} \label{conf rule 1}
&\l \Phi_k(x) \r = 0, \qquad \qquad \qquad  k\neq 0, \\
&\l \Phi_{k_1}(x_1) \Phi_{k_2}(x_2) \r = 0, \qquad k_1 \neq k_2.
\label{conf rule 2}
\end{align}

For higher correlation numbers we use the OPE to bring the $n$-point correlation function to a combination of two-point functions and then use the rules (\ref{conf rule 1}),(\ref{conf rule 2}). For instance the three-point correlation functions satisfy

\be \label{conf rule 3}
\l \Phi_{k_1} \Phi_{k_2} \Phi_{k_3} \r = 0 , \qquad k_3 > I(k_1,k_2) = \begin{cases} k_1+k_2, \qquad\qquad k_1+k_2 \leq p-2 \\ 2p-k_1-k_2 -4, \quad k_1+k_2>p-2 \end{cases}
\ee
where we assume that $k_1 \leq k_2 \leq k_3$. For four-point correlation functions we have

\be \label{conf rule 4}
\l \Phi_{k_1} \Phi_{k_2} \Phi_{k_3} \Phi_{k_4} \r = 0 , \qquad k_4 > I(I(k_1,k_2),k_3)
\ee
where we assume that $k_1 \leq k_2 \leq k_3 \leq k_4$ and

\be 
I(I(k_1,k_2),k_3) = \begin{cases} k_1+k_2+k_3; \qquad \qquad \quad\qquad k_1+k_2 \leq p-2, \quad k_1+k_2+k_3 \leq p-2 \\ 
2(p-2)-k_1-k_2+k_3; \qquad k_1+k_2>p-2, \quad k_1+k_2-k_3 \geq p-2 \\
2(p-2)-k_1-k_2-k_3; \qquad k_1+k_2 \leq p-2, \quad k_1+k_2+k_3 >p-2 \\
k_1+k_2-k_3; \qquad\qquad\quad\qquad k_1+k_2 >p-2, \quad k_1+k_2 -k_3<p-2 \end{cases}
\ee

The equations like (\ref{conf rule 1})-(\ref{conf rule 4}) we call \emph{selection rules}.


\subsection{Coupling to Liouville theory}
${}$

In this section we consider the interaction of Minimal Models, considered in the previous section, with the Liouville filed theory. The Minimal Model plays the role of conformal matter. For this reason the resulting theory is called Minimal Liouville Gravity.

The Polyakov's continuous approach to two-dimensional quantum gravity \cite{Polyakov:1981rd} is defined through the path integral over two-dimensional Riemannian metrics $g_{\mu\nu}$ interacting with some conformal matter. In conformal gauge $g_{\mu\nu} = e^{\phi} \hat{g}_{\mu\nu}$ it leads to Liouville action

\be
S_L = {1\over 4\pi} \int_M \sqrt{\hat{g}} \left( \hat{g}^{\mu\nu} \partial_{\mu}\phi\partial_{\nu}\phi + Q \hat{R} \phi +4 \pi \mu e^{2b\phi} \right)d^2x
\ee  
where $\hat{g}_{\mu\nu}$ is some fixed background metric, $\mu$ -- cosmological constant and parameters $Q, b$ are related to the central charge $c_L$ of the Liouville theory 

\be 
c_L = 1+6Q^2, \qquad Q = b+b^{-1}.
\ee
 
The central charge $c_M$ of the conformal matter is related to the central charge of the Liouville theory by the Weyl anomaly cancellation condition

\be 
c_L + c_M = 26.
\ee

In the case of $(q,p)$ Minimal Liouville Gravity where the conformal matter is $(q,p)$ Minimal Model of CFT, we have $b = \sqrt{q\over p}$.

The observables of the $(q,p)$ Minimal Liouville Gravity are constructed as cohomologies of an appropriate BRST operator. They are enumerated by the same integers as primary fields in corresponding Minimal Model and denoted as $O_{m,n}$. Explicitly they have the form 

\be \label{observables}
O_{m,n} = \int_{x \in M} {\cal O}_{m,n}(x), \qquad  {\cal O}_{m,n}(x) = \Phi_{m,n}(x)e^{2b \delta_{m,n}\phi(x)}\sqrt{\hat{g}}d^2x
\ee 
where $\Phi_{m,n}$ -- primary fields of the $(q,p)$ CFT Minimal Model, considered in the previous section. The numbers $\delta_{m.n}$ are the so-called gravitational dimensions \cite{Knizhnik:1988ak}

\be \label{gravitaional dimensions}
\delta_{m,n} = {p+q-|pm-qn| \over 2q}.
\ee
The operators $O_{m,n}$ have the scaling property 

\be \label{observable's dimension}
O_{m,n} \sim \mu^{-\delta_{m,n}}.
\ee
Obviously, the operators $O_{m,n}$ satisfy the same selection rules (\ref{conf rule 1}) -- (\ref{conf rule 4}) as the primary fields $\Phi_{m,n}$ do. 

The most easily defined physical object in Minimal Liouville Gravity is correlation numbers

\be \label{correlation numb}
 Z_{m_1n_1\dots m_Nn_N} = \l O_{m_1,n_1} \dots O_{m_N,n_N} \r
\ee
where the average is taken over the fluctuations of the metric and of the conformal matter fields. Again it is convenient to consider the generating function of these correlation numbers

\be \label{gen func}
Z_L(\lambda) =\left\langle \exp  \sum_{m,n} \lambda_{m,n} O_{m,n}\right\rangle.
\ee
We can also think about the generating function as the partition function of the Minimal Gravity perturbed by the operators $O_{m,n}$ with the coupling constants $\lambda_{m,n}$. For that reason the generating function will often be called \emph{partition function}.

This kind of generating functions and correlation numbers will be the main object of our study. In particular we will check that the results of \cite{Belavin:2006ex} for the correlation numbers in $(q,p)$ Minimal Gravity agree with the those obtained from the so-called Douglas string equation, which will be introduced in the subsequent sections.


\subsection{Contact terms}

In this section we discuss one important subtlety which makes the definition of the correlation numbers ambiguous. The correlation numbers (\ref{correlation numb}) involve integration over $n$ points on the two-dimensional surface $M$ 

\be 
Z_{m_1n_1\dots m_Nn_N} = \int \l {\cal O}_{m_1,n_1}(x_1) \dots {\cal O}_{m_N,n_N}(x_N) \r d^2x_1 \dots d^2x_N
\ee

Let us notice that there could be contact delta-like terms when two or more points $x_i$ are coincident. Such terms are not controlled by the conformal field theory and thus make the definition of correlation numbers ambiguous. For instance, in CFT calculations, like those in \cite{Belavin:2006ex}, one just disregards all the contact terms and supposes that the conformal selection rules are the same for operators $O_{m,n}$ and ${\cal O}_{m,n}$. On the other hand in the discrete approach, and in stemming from it Douglas string equation, one might expect that the contact terms are treated differently and conformal selection rules for correlators of $O_{m,n}$ are not satisfied. This is what actually happens. Apparently this phenomenon was first observed in \cite{Moore:1991ir}.

In other words the ambiguity in contact terms leads to the fact that we can add to the $n$-point correlation numbers some $k$-point correlation numbers, where $k<n$. For instance, we can add a one-point correlation number to the two-point correlation number 

\be 
\l O_{m_1,n_1} O_{m_2,n_2} \r \to \l O_{m_1,n_1}O_{m_2,n_2} \r + \sum_{m,n} A_{m,n}^{(m_1n_1)(m_2n_2)} \l O_{m,n} \r.
\ee
It is easy to see that such a substitution is equivalent to a change of coupling constants in the generating function (\ref{gen func})

\be \label{introduction resonance relations}
\lambda_{m,n} \to \lambda_{m,n} + \sum_{m_1,n_1,m_2,n_2} A_{m,n}^{(m_1n_1)(m_2n_2)} \lambda_{m_1n_1}\lambda_{m_2n_2}.
\ee 
Also we must include here possible changes of the background \cite{Moore:1991ir} if $\lambda_{m,n}$ is proportional to an integer power of the cosmological constant $\mu = \lambda_{1,1}$. Thus we arrive at a convenient way of describing the ambiguity in contact terms. Namely, any addition of contact terms is equivalent to some non-linear polynomial change of coupling constants in the generating function (\ref{gen func})
 \begin{align}
\label{introduction resonance relations 2}
\lambda_{m,n} \to C_{m,n}\mu^{\delta_{m,n}} + \lambda_{m,n} &+ \sum_{m_1,n_1}\sum_{m_2,n_2} C_{m,n}^{(m_1n_1)(m_2n_2)} \lambda_{m_1n_1}\lambda_{m_2n_2} + \nn \\
&+\sum_{m_1,n_1}\sum_{m_2,n_2}\sum_{m_3,n_3} C_{m,n}^{(m_1n_1)(m_2n_2)(m_3n_3)} \lambda_{m_1n_1}\lambda_{m_2n_2}\lambda_{m_3n_3}+ \dots .
\end{align}
So, there can be many different ``systems of coordinates" of $\lambda_{m,n}$. In Minimal Gravity we use one ``system of coordinates" and the Douglas string equation uses another one. In Minimal Gravity there is a strong restriction on the substitutions of coupling constants (\ref{introduction resonance relations 2}).
Indeed, the observables $O_{m,n}$ have certain mass dimension (\ref{observable's dimension}). Consequently the coupling constants also have certain mass dimension
\be \label{coupling constants dim}
\lambda_{m,n} \sim \mu^{\delta_{m,n}}
\ee
This restricts the possible non-linear terms in (\ref{introduction resonance relations 2}). Namely, all the acceptable terms in (\ref{introduction resonance relations 2}) must have the same dimension as $\lambda_{m,n}$
\be \label{resonance condition}
\delta_{m,n} = \delta_{m_1,n_1} + \delta_{m_2,n_2} + \delta_{m_3,n_3} + \dots 
\ee 
We usually call (\ref{resonance condition})  \emph{the resonance condition} and substitutions of coupling constants (\ref{introduction resonance relations 2}) \emph{resonance relations}. 

In the Liouville Minimal Gravity the resonance conditions (\ref{resonance condition}) turn out to be very common. So we must resolve this ambiguity somehow. We do it in the following way. We evaluate the free energy ${\cal F}(t)$ from the Douglas equation. Then make a change of variables $t=t(\lambda)$ like in (\ref{introduction resonance relations 2}) and require that after this substitution the conformal selection rules are satisfied.


\subsection{Three- and four-point correlation numbers}

In \cite{Goulian:1990qr}, \cite{Zamolodchikov:2005sj} the three-point correlation numbers were calculated. The four-point correlation numbers were calculated in \cite{Belavin:2006ex}. We will compare these results with those from the Douglas string equation. 

A priori, the normalizations of the operators $O_{m,n}$ and of the correlators are not coincident in Minimal Gravity and in the Douglas string equation. Thus to make sensible comparisons we write down the quantities which do not depend on the normalizations of operators and correlators  

\begin{align} \label{3-point MG}
{ \l\l O_{m_1,n_1}O_{m_2,n_2}O_{m_3,n_3} \r\r^2 \over \prod_{i=1}^3 \l\l O^2_{m_i,n_i}\r\r }&=  {\prod_{i=1}^3 |m_ip - n_iq| \over p(p+q)(p-q)} \\
{ \l\l O_{m_1,n_1}O_{m_2,n_2}O_{m_3,n_3}O_{m_4,n_4} \r\r \over \left( \prod_{i=1}^4 \l\l O^2_{m_i,n_i}\r\r\right)^{1\over 2} } &= \nn \\ = {\prod_{i=1}^4 |m_ip-n_iq|^{1\over 2} \over 2p(p+q)(p-q)}& \left( \sum_{i=2}^4 \sum_{r=-(m_1-1)}^{m_1-1}\sum_{t=-(n_1-1)}^{n_1-1} |(m_i-r)p-(n_i-t)q| -m_1n_1(m_1p+n_1q) \right)
\label{4-point MG}
\end{align}
where $\l\l \dots \r\r = {\l \dots \r \over \l 1 \r}$. Sums over $r,t$ in the last formula are with the step 2. 

Additionally it needs to be mentioned that the four-point correlation numbers (\ref{4-point MG}) were obtained under certain assumptions. In the particular case when $m_i =1, i=1,\dots ,4$ the assumption looks as follows (cf. \cite{Belavin:2006ex}) 
\be \label{4-point condition}
n_1+n_4 \leq n_2+n_3.
\ee 
(suppose $n_1 \leq \dots \leq n_4$).


\section{Douglas string equation approach to Minimal Gravity}

\subsection{Douglas string equation}

Due to Douglas \cite{Douglas:1989dd} the free energy of the matrix model, corresponding to the partition function (\ref{gen func}) of $(q,p)$ Minimal Gravity, is described in the following way. One takes differential operators $P$ and  $Q$ of the form

\begin{align} \label{Q def}
\hat{Q} &= d^q+\sum_{\alpha =1}^{q-1}u_{\alpha}(x)d^{q-\alpha -1} ,\\
\label{P def}
\hat{P} &= \left({1\over q}\left(1+{p\over q}\right) \hat{Q}^{p\over q} + \sum_{\alpha=1}^{q-1}\sum_{k=1}  {1\over q}\left(k+{\alpha\over q}\right) t_{k,\alpha} \hat{Q}^{k+ {\alpha \over q}-1} \right)_+
\end{align}
where $d={d\over dx}$, the pseudo-differential operators $\hat{Q}^{a\over q}=d^a+\dots$ are understood as series in $d$ and $(\dots)_+$ means that only non-negative powers of $d$ are taken in the series expansion over $d$. The constants $t_{k,\alpha}$ are usually called ``times". This terminology comes from the connection with KdV-type integrable hierarchies (see Appendix A below; the constants $t_{k,\alpha}$ coincide, up to a numerical factor, with the time variables $t^\alpha_k$ of the Gelfand--Dickey hierarchy).

Then the so-called \emph{string equation} is defined as

\begin{align} 
[\hat{P}, \hat{Q}] = 1 \label{Douglas eq}.
\end{align}
The string equation yields a system of differential equations on $u_{\alpha}(x)$. The free energy ${\cal F}(\tilde t)$ then satisfies the equation 

\be  \label{douglas free energy}
{\partial^2 {\cal F} \over \partial x^2} = u_1^*
\ee
where $u_{\alpha}^*$ is an appropriate solution of the string equation (\ref{Douglas eq}).

The first term in (\ref{P def}) describes the critical point of the matrix model and corresponds to the Minimal Liouville Gravity. Other terms in (\ref{P def}) correspond to the Minimal Liouville Gravity perturbed by primary operators. In (\ref{P def}) we did not indicate the interval in which the index $k$ is varied. We will do this later from the condition that the perturbations by the times $t_{k,\alpha}$ correspond to the coupling constants $\lambda_{m,n}$ of the perturbation of the Minimal Liouville Gravity by all primary operators from $(q,p)$ minimal CFT. 

In the present work we are interested in calculating the correlation numbers on the sphere. In the language of Douglas string equation this means that we are interested in the quasiclassical (dispersionless) limit of these equations. In this limit the ``momentum operator" $d\over dx$ is replaced with a variable\footnote{Not to be confused with $p$ in the $(q,p)$ minimal CFT.} $p$  and the commutator in string equation is replaced with Poisson bracket 

\begin{align}
\label{quasiclassical string equation}
[P,Q] &=1 \to \{ P, Q\} = {\partial P \over \partial x}{\partial Q \over \partial p} -{\partial P \over \partial p} {\partial Q \over \partial x}=1, \\
Q &= p^q + \sum_{\alpha =1}^{q-1}u_{\alpha}(x) p^{q-\alpha -1}, \label{Q polynomial} \\
P &= \left( {1\over q}\left(1+{p\over q}\right) Q^{p\over q} + \sum_{\alpha=1}^{q-1}\sum_{k=1} {1\over q}\left(k+{\alpha\over q}\right) t_{k,\alpha} Q^{k+ {\alpha \over q}-1} \right)_+ 
\end{align}
where $P$ and $Q$ are the symbols of the differential operators $\hat{P}$ and $\hat{Q}$. $Q^{a\over q}$ is understood as series expansion in $p$ and $(\dots)_+$ means that only non-negative powers of $p$ are taken in this expansion.

The string equation (\ref{Douglas eq}) is actually a system of differential equations on the functions $u_{\alpha}$. Heuristically, taking the  quasiclassical limit means that we must through out all the derivative terms except the first derivatives (leave terms linear in $du_{\alpha} \over dx$ and get rid of terms ${d^2u_{\alpha} \over dx^2}, {d^3u_{\alpha} \over dx^3}, \dots$). 

It was shown in \cite{Ginsparg:1990zc} that either in the full form (\ref{Douglas eq}) or the quasiclassical limit (\ref{quasiclassical string equation}), one can take the first integral of the string equation. In the quasiclassical limit after this integration the Douglas string equation (\ref{Douglas eq}) becomes \cite{Ginsparg:1990zc}  (see the proof in Appendix A)

\be \label{action principle}
{\p S \over \p u_\alpha(x)} = 0
\ee
where

\be \label{action ambigious}
S[u_{\alpha}(x)]  = S_{s+1,p_0} +  \sum_{\alpha =1}^{q-1}\sum_{k=0} t_{k,\alpha} S_{k,\alpha}
\ee
where $(q,p)=(q,sq+p_0)$, $1\leq p_0\leq q-1$, $t_{0,1} =x$ and we introduced a notation
\be
S_{k,\alpha}  =  Res (Q^{k+ {\alpha \over q}}) \label{hamiltonians}
\ee
Here $Q$ is the polynomial (\ref{Q polynomial}) and the residue is taken at $p=\infty$. The details can be found in the Appendix A.
\footnote{The functions $S_{k,\alpha}$ are the same as the functions $\theta_{\alpha,k}$ in the Appendix A up to some numerical factor.}

The times $t_{0,\alpha}$ in (\ref{action ambigious}) with $2\leq \alpha \leq q-1$, which do not appear in (\ref{P def}), are the integration constants. The time $t_{0,1}$, which does not appear in (\ref{P def}) either, comes from the integration of the unit on the right hand side in (\ref{Douglas eq}).

For obvious reasons, the equation (\ref{action principle}) is usually referred to as the principle of least action. Since it is equivalent to the Douglas string equation, we will be using the former for convenience. For instance, as was mentioned above, we are interested in the quasiclassical limit of the string equation. For the string equation itself it means that we are left with the system of the first order differential equations. On the other hand for the principle of least action equation (\ref{action principle}) it means that in the quasiclassical limit we are left with the system of algebraic equations. So it is convenient for us to use the principle of least action instead of the Douglas string equation.

Now let us make more accurate statement about the perturbations we need to be added in (\ref{action ambigious}). Namely, we ask in what region we should vary the index $k$ in (\ref{action ambigious}) in order to make the correspondence with the perturbation by all primary operators in Minimal Liouville Gravity. This question can be answered with the help of dimensional analysis. The construction of the Douglas string equation happens to have a scaling properties similar to those of the Minimal Liouville Gravity. Historically, this coincidence of the ``gravitational dimensions" was actually the first reason to believe that two descriptions describe the same phenomena.

To identify the free energy of the matrix model with the partition function of the Liouville Minimal Gravity on the sphere, we first analyse the dimensions. 
The $(q,p)$ Liouville Minimal Gravity partition function on the sphere has the following scaling property \cite{Knizhnik:1988ak}

\be Z_L \sim \mu^{{p+q\over q}} \ee
where $\mu$ is the cosmological constant.

On the other hand the equation (\ref{douglas free energy}) gives that

\be Z \sim x^2 u_1 \sim p^{2(p+q)} \ee
where the scaling dimension of $x$ is determined from the equation $\{P,Q\}=1$ and scaling dimension of $u^{\alpha}$ is determined from the constraint that all terms in the polynomial $Q$ are of the same order

\be x \sim p^{p+q-1}, \qquad u_{\alpha} \sim p^{\alpha +1}. \ee 
Thus if we want ${\cal F} \sim Z_L$ we have 

\be p\sim \mu^{1\over 2q}.\ee 
This determines 

\be \label{t dimensions}
t_{k,\alpha} \sim \mu^{{s+1-k\over 2}+{p_0 - \alpha \over 2q}}.
\ee
In particular $t_{s-1, p_0} \sim \mu$.
We want to identify these times and their dimensions with the dimensions of the coupling constants $\lambda_{m,n}$ (\ref{coupling constants dim}) or, equivalently, with the dimensions of the observables $O_{m,n}$ in the Minimal Liouville Gravity (\ref{observable's dimension}).

In the $(q,p)$ Minimal Gravity one has the following scaling dimensions of the coupling constants $\lambda_{m,n}$ ($1 \leq m \leq q-1; 1\leq n \leq p-1$)(\ref{gravitaional dimensions}), see eqs. (\ref{coupling constants dim}) above

\be \label{grav dim}
\delta_{m,n} =  {p+q -|pm-qn| \over 2q}.
\ee

Let us find among the times $t_{k,\alpha}$ those which correspond to the coupling constant $\lambda_{m,n}$ in Minimal Gravity. To do this observe that the expression for the ``action" (\ref{action ambigious}) determines the dimensions of the times $t_{k,\alpha}$. Let us write it in the form $S = Res\left( Q^{p+q \over q} + \sum_{m,n} \tau_{m,n}Q^{a_{m,n}} \right)$, where we changed enumeration of times $(k,\alpha)$ to some new enumeration $(m,n)$. Then the dimension of times $\tau_{m,n}$ is (since $Q \sim p^q \sim \mu^{1\over 2}$)

\be 
\tau_{m,n} \sim \mu^{{1\over 2}\left({p+q \over q} - a_{m,n} \right)  }.
\ee 

Now we see that in order for the dimension of the time $\tau_{m,n}$ to coincide with (\ref{grav dim}), we need to put $a_{m,n} = {|pm-qn| \over q}$. Thus we have another expression for the action, in which the correspondence with the operators from Minimal Gravity is clear

\be \label{action unambiguous}
S = Res \left( Q^{p+q \over q} + \sum_{m = 1}^{q-1} \sum_{n =1}^{p-1} \tau_{m,n}Q^{|pm-qn| \over q} \right).
\ee

The relation between two types of enumeration $(k,\alpha)$ and $(m,n)$ can be established from the equation 

\be 
{|pm-qn| \over q} = k+ {\alpha \over q}.
\ee
For instance $\mu \sim t_{s-1,p_0} = \tau_{1,1}$, where $(q,p) = (q,sq+p_0)$, $1\leq p_0 \leq q-1$.

So far we have seen that the dimensions of times $\tau_{m,n}$ coincide with the dimensions of the coupling constants $\lambda_{m,n}$. However, as it was mentioned in the section 2.3, they do not necessarily coincide but can have a non-linear relation like in (\ref{introduction resonance relations 2})

\begin{align} \label{resonance substitution}
\tau_{m,n} = C_{m,n}\mu^{\delta_{m,n}}+  \lambda_{m,n} &+ \sum_{m_1,n_1} C_{m,n}^{(m_1n_1)} \lambda_{m_1n_1}\mu^{\delta_{m,n}-\delta_{m_1,n_1}} + \nn \\
&+\sum_{m_1,n_1}\sum_{m_2,n_2} C_{m,n}^{(m_1n_1)(m_2n_2)} \lambda_{m_1n_1}\lambda_{m_2n_2}\mu^{\delta_{m,n}-\delta_{m_1,n_1}-\delta_{m_2,n_2}}+ \dots .
\end{align}
There is a strong restriction on the possible terms, namely, they must have the same scaling dimension as $\lambda_{m,n}$ and the terms on the r.h.s are non-zero only when the cosmological constant $\mu$ appears in a non-negative integer power. These relations for the series of $(2,2s+1)$ Minimal Gravities were found in \cite{Belavin:2008kv}. We are aiming to find them in the case of $(3,3s+p_0)$ Minimal Gravity. We will do it after introducing an explicit expression for the partition function which satisfies the equation (\ref{douglas free energy}).

\subsection{Free energy in the Douglas approach}

The free energy ${\cal F}$ in the Douglas approach satisfies the differential equation 

\be \label{partition function equation}
{\partial^2 {\cal F} \over \partial x^2} = u_1^* 
\ee
where $u_1^*$ is the solution of Douglas string equation (\ref{quasiclassical string equation}) or, equivalently, (\ref{action principle}).

Nevertheless it is desirable to have an explicit expression for the partition function. Such an expression exists due to relation of the partition function satisfying the equation (\ref{partition function equation}) to the Sato's tau function \cite{Segal1985}. Let us introduce necessary definitions to write down the explicit formula for the partition function satisfying the equation (\ref{partition function equation}). (This is deeply related to Frobenius manifolds. More details can be found in Appendix A.)

First of all introduce an algebra of polynomials modulo the polynomial $Q'$, where prime denotes the derivative over $p$

\be \label{algebra}
{\cal A} = C[p]/Q'. 
\ee
Also define a bilinear form on the space of such polynomials

\be 
(P_1(p),P_2(p)) :=Res \left({P_1(p)P_2(p) \over Q'}\right),
\ee
where, as before, the residue is taken at $p=\infty$. Define multiplication on the space ${\cal A}$ by introducing some basis $\phi_{\alpha}$ and define

\be 
\phi_{\alpha} \phi_{\beta} = C_{\alpha \beta}^{\gamma} \phi_{\gamma} \mod(Q').
\ee 
Besides we always take the unity as the first element of the basis: $\phi_1 = 1$. Thus we have $C^{\alpha}_{1\beta} = \delta^{\alpha}_{\beta}$.

One then finds that 

\be 
Res {\phi_{\alpha}\phi_{\beta}\phi_{\gamma} \over Q'} =  C_{\alpha \beta}^{\delta} \cdot Res {\phi_{\delta}\phi_{\gamma} \over Q'} = 
C_{\alpha \beta}^{\delta}g_{\delta\gamma} =                 C_{\alpha\beta\gamma}
\ee
where the indices are raised and lowered by the metric $g_{\alpha\beta}$ 

\be \label{metric}
g_{\alpha\beta} = (\phi_{\alpha}, \phi_{\beta}).
\ee

There are of course many possible choices for the basis $\phi_{\alpha}$. However there are two particularly useful for us. The first one is just to take as a basis the monomials $p^{\alpha}$. 

Besides, there is another useful basis, namely 

\be \phi_{\alpha} = {\partial Q \over \partial v^{\alpha}}
\ee
where 
\be 
\qquad v^{\alpha} = - {q\over q-\alpha} Res Q^{q-\alpha \over q}. \ee
This basis has a nice property that the metric in this basis takes a form $g_{\alpha\beta} = \delta_{\alpha+\beta,q}$ (see Appendix A). Using (\ref{algebra}) - (\ref{metric}) one can prove that in this basis \cite{Dijkgraaf:1990dj}

\be \label{DVV relation}
C_{\alpha\beta\gamma} = {\p^3 F \over \p v^{\alpha} \p v^{\beta} \p v^{\gamma} }.
\ee
where the structure constants $C_{\alpha\beta\gamma}$ are evaluated in the basis ${\partial Q \over \partial v^{\alpha}}$.
The proof of this relation can be found in \cite{Dijkgraaf:1990dj} and also in the Appendix A.
The formulae (\ref{DVV relation}) together with (\ref{algebra}) - (\ref{metric}) gives the structure of the Frobenius manifold on ${\cal A}$ \cite{Dubrovin:1992dz}, $v^{\alpha}$ being the flat coordinates on this manifold.

The claim is that the free energy can be written as follows

\be \label{free energy integral}
{\cal F} = \frac12\int_0^{u_*} C^{\beta\gamma}_{\alpha} {\partial S \over \partial u^{\beta}} {\partial S \over \partial u^{\gamma}} du^{\alpha}
\ee
where $u_*$ is an appropriate solution of Douglas string equation or equivalently an appropriate critical point of the action $S$. Here the structure constants $C^{\beta\gamma}_{\alpha}$ are evaluated in the basis $\phi_{\alpha} = p^{\alpha}$. The indices are taken up or down by the metric $g_{\alpha\beta}$. 

To make sure that this formula for the partition function makes sense, we need to verify two points. 

The first is that the free energy ${\cal F}$ does not depend on the contour of integration. Equivalently,  the differential one-form $\Omega = C_{\alpha}^{\beta\gamma} {\partial S \over \partial v^{\beta}} {\partial S \over \partial v^{\gamma}} dv^{\alpha}$ must be closed. It is convenient to use the flat coordinates $v^{\alpha}$. This is checked by direct calculation of de Rham differential of $\Omega$ and using two properties, namely, associativity of the algebra $\cal A$ and the recursion relation for hamiltonians $S_{k,\alpha}$

\begin{align}
C_{\alpha\beta}^{\gamma}C_{\gamma \delta}^{\phi} &= C_{\alpha\gamma}^{\phi}C_{\beta\delta}^{\gamma}, \\
{\partial^2 \theta_{\alpha,n} \over \partial v^{\beta} \partial v^{\gamma}} &= C_{\beta\gamma}^{\delta} {\partial \theta_{\alpha,n-1} \over \partial v^{\delta}}
\end{align} 
where $\theta_{\alpha,n}$ is the same function as $S_{n,\alpha}$ up to a normalization factor which is defined in the Appendix A. The proof of the recursion relation is also provided in the Appendix A.

The second is that the partition function (\ref{free energy integral}) must satisfy the equation (\ref{partition function equation}). This is easily verified by direct differentiation of (\ref{free energy integral}) and using that $t_{0,1}=x$ and $C_{\alpha 1}^{\beta} = \delta_{\alpha}^{\beta}$.

Now we have a nice explicit expression (\ref{free energy integral}) for the partition function which is convenient to use. In order to calculate the correlation numbers we make the substitution (\ref{resonance substitution}) in the partition function(\ref{free energy integral}) and then just take derivatives

\be \label{correlation numbers der}
Z_{m_1n_1 \dots m_Nn_N} = {\partial \over \partial \lambda_{m_1n_1}} \dots {\partial \over \partial \lambda_{m_Nn_N} } \Bigg|_{\substack{\lambda_{m,n} =0 \\ \text{for }(m,n) \neq (1,1)}} {\cal F}[t(\lambda)].
\ee 
where after taking derivatives we take $\lambda_{m,n}=0$ except for the cosmological constant $\mu = \lambda_{1,1}$. 

Now we are going to consider explicitly examples of $(2,2s+1)$ and $(3,3s+p_0)$ Minimal Gravity. The former was considered in \cite{Belavin:2008kv}, however we rederive the results of that paper with a few details changed. This will allow us to make some generalizations to the case $(3,3s+p_0)$.


\section{${\bf (q,p) = (2,2s+1)}$ Minimal Gravity}

In this case we have the polynomial $Q$ (\ref{Q polynomial}) and the free energy (\ref{free energy integral})

\be \label{partition function 2s}
Q = p^2 + u, \qquad {\cal F} = {1\over 2}\int_0^{u_*} S_u^2(u)du
\ee
where $u_*$ is the solution of the equation 

\be 
S_u \equiv {\p S \over \p u} = u^{s+1} + \sum_{n=1}^s \tau_{1,n} u^{s-n} = 0
\ee
where for simplicity we have changed the normalization of the times $\tau_{m,n}$ and overall normalization of the action (\ref{action unambiguous}).

To get the generating function for the correlation numbers one needs the resonance relations 
\footnote{It is convenient to make a shift of indices here in order to single out the cosmological constant $\mu$ since it plays a special role in our discussions. So $\lambda_0 \sim \tau_{1,1} \sim \mu$ corresponds to the operator $O_{1,1}$.  }

\be \label{res 2,2s+1}
{\tau}_{1,k+1} = \lambda_k + C_k \mu^{\delta_k} + \sum_{l,k_1} C_k^{lk_1} \mu^l \lambda_{k_1} + \sum_{l,k_1,k_2}  C_k^{lk_1k_2}\mu^{l}\lambda_{k_1}\lambda_{k_2} +  \dots \sum_{l,k_1,\dots, k_n}  C_k^{lk_1 \dots k_n}\mu^{l} \lambda_{k_1}\dots \lambda_{k_n} + \dots \ee
where all the powers of $\mu$ should appear with non-negative integer powers. In Liouville Minimal Gravity these relations always contain only finite number of terms. This due to the restriction that all the terms must be of the same gravitational dimension.

Now we can insert the resonance relations (\ref{res 2,2s+1}) into the partition function and polynomial $S_u$. The result is of the form

\begin{align}
{\cal F} &= Z_0 + \sum_{k=1}^{s-1} \lambda_{k} Z_k + {1\over 2} \sum_{k_1,k_2=1}^{s-1} \lambda_{k_1} \lambda_{k_2} Z_{k_1k_2} + \dots  \\
S_u &= S_u^0 + \sum_{k=1}^{s-1} \lambda_{k} S_u^k + {1\over 2} \sum_{k_1,k_2=1}^{s-1} \lambda_{k_1} \lambda_{k_2} S_u^{k_1k_2} + \dots  \\
\end{align}
Also from the original form of the polynomial $S_u$ one finds that

\begin{align}
S_u^0(u) &= u^{s+1} + B\mu u^{s-1} + C \mu^2 u^{s-3} + \dots \\
S_u^k(u) &= A_k u^{s-k-1} + B_k \mu u^{s-k-3} + C_k \mu^2 u^{s-k-5} +\dots \\
S_u^{k_1k_2}(u) &= A_{k_1k_2} u^{s-k_1-k_2-3} + B_{k_1k_2} \mu u^{s-k_1-k_2-5} + C_{k_1k_2} \mu^2 u^{s-k_1-k_2-k_3-7} + \dots \\
\dots \nn
\end{align}
where all the polynomials have certain parity since $\mu \sim u^2$.

From the discussion of section 3.1 we find that the dimensions in this case are

\begin{align}
\lambda_k &\sim \mu^{k+2\over 2} \\
{\cal F} &\sim \mu^{2s+3\over 2} \\
Z_{k_1\dots k_n} &\sim \mu^{2s+3-\sum(k_i + 2) \over 2} \label{2s dim corr numb}
\end{align}

As usually in the spirit of the scaling theory of criticality, we are interested only in the singular part of the partition function and disregard the regular part as non-universal. 

Notice that $Z_{k_1 \dots k_n}$ is always singular if $\sum k_i$ is even. On the other hand when $\sum k_i$ is odd and additionally 

\be \sum_{i=1}^n k_i \leq 2s+3-2n \ee 
the correlation number $Z_{k_1 \dots k_n}$ involves only non-negative integer powers of $\mu$ and thus is non-singular. This inequality always holds for one- and two-point correlation numbers. So we shall consider the sector of odd $\sum k_i$ only starting from the three point correlation numbers.

\subsection{Dimensionless setup}
As in \cite{Belavin:2008kv} it is convenient to switch to dimensionless quantities

\begin{align}
s_k &= {g_k \over g} u_0^{-(k+2)} \lambda_k \\
S_u(u) &= g u_0^{s+1} Y_u({u/ u_0}) \\
{\cal F} &= g^2 u_0^{2s+3} {\cal Z}
\end{align}
where $u_0 = u_*(\lambda=0) \sim \mu^{1\over 2}$, $g_k = {(p-k-1)! \over (2p-2k-3)!!}$ and $g={(p+1)! \over (2p+1)!!}$\footnote{For details see \cite{Belavin:2008kv}. Our polynomial $Y_u$ is denoted as $Q$ there.}. 

Then one has

\be
{\cal Z} = {1\over 2} \int_0^{x_*} Y_u^2(x) dx 
\ee
where $x={u\over u_0}$ and $x_* = x_*(s)$ is an appropriate zero of the polynomial $Y_u(x)$. Notice that $x_*(s=0)=1$. 

Similarly to dimensional quantities one has expansions

\begin{align}
{\cal Z} &= {\cal Z}_0 + \sum_{k=1}^{s-1} s_{k} {\cal Z}_k + {1\over 2} \sum_{k_1,k_2=1}^{s-1} s_{k_1} s_{k_2} {\cal Z}_{k_1k_2} + \dots  \\
Y_u&= Y_u^0 + \sum_{k=1}^{s-1} s_{k} Y_u^k + {1\over 2} \sum_{k_1,k_2=1}^{s-1} s_{k_1} s_{k_2} Y_u^{k_1k_2} + \dots \\
Y_u^0(x) &= C_0 x^{s+1} + C_0'  x^{s-1} + \dots \label{Q0 exp} \\
Y_u^k(x) &= C_k x^{s-k-1} + C_k' x^{s-k-3} +\dots \label{Qk exp} \\
Y_u^{k_1k_2}(x) &= C_{k_1k_2} x^{s-k_1-k_2-3} + C_{k_1k_2}' x^{s-k_1-k_2-5} +\dots \label{Qkk exp}  \\
\dots \nn
\end{align}

\subsection{One- and two-point correlation numbers}
Using (\ref{correlation numbers der}) and the partition function (\ref{partition function 2s}) we find one- and two-point correlation numbers 

\begin{align} \label{one point}
{\cal Z}_k &= \int_0^1 dx Y_u^0(x) Y_u^k(x)  \\
{\cal Z}_{k_1k_2} &= \int_0^1 dx (Y_u^{k_1}(x) Y_u^{k_2}(x) + Y_u^0(x)Y_u^{k_1k_2}(x)) \label{two point}
\end{align}
where, due to (\ref{2s dim corr numb}), the one-point correlation numbers are singular only for even $k$ and the two-point correlation numbers are singular only for even $k_1+k_2$.

Fusion rules (\ref{conf rule 1}), (\ref{conf rule 2}) demand that one-point correlation numbers are zero for $k\neq 0$ and two point numbers are zero when $k_1 \neq k_2$. The second term in the two point numbers is actually absent due to (\ref{one point}) and to the fact that the polynomial $Y_u^{k_1k_2}$ can be written as a linear combination of $Y_u^k$. 

Also it is convenient here to introduce a new variable $y$ instead of $x$

\be \label{change variable}
{y+1 \over 2} = x^2, \qquad dx = {dy \over 2 \sqrt{2} (1+y)^{1\over 2}} 
\ee

In terms of the variable $y$ the polynomials
\footnote{For simplicity we denote the polynomial $Y(x(y))$ just as $Y(y)$.}
$Y_u^0, Y_u^k, \dots$ will contain all powers instead of going with step 2 (\ref{Q0 exp}), (\ref{Qk exp}). And the shift was made in order for the interval of integration in (\ref{one point}), (\ref{two point}) to be $[-1,1]$ instead of $[0,1]$. 

Thus the fusion rules condition become an orthogonality condition on the polynomials $Y_u^0, Y_u^k$

\begin{align} 
{\cal Z}_k &= \int_{-1}^1 {dy \over 2 \sqrt{2} (1+y)^{1\over 2}} Y_u^0(x) Y_u^k(x) = 0, \quad k \neq 0 \\
{\cal Z}_{k_1k_2} &= \int_{-1}^1 {dy \over 2 \sqrt{2} (1+y)^{1\over 2}} Y_u^{k_1}(x) Y_u^{k_2}(x) =0, \quad k_1\neq k_2
\end{align}
and, together with the condition $Y_u^0(1)=0$, it determines the polynomials $Y_u^0$ and $Y_u^k$:
\vspace {0.4cm}
\begin{center}
\begin{tabular}{|c|c|}
\hline 
s~ odd & $Y_u^0(y) = P_{s+1\over 2}^{(0,-{1\over 2})}(y) - P_{s ~1\over 2}^{(0,-{1\over 2})}(y)$   \\ 
s ~ even & $Y_u^0(y) = x \left( P_{s\over 2}^{(0,{1\over 2})}(y) - P_{s-2\over 2}^{(0,{1\over 2})}(y) \right)$
\\
s+k ~ odd & $Y_u^k(y) = P_{s-k-1\over 2}^{(0,-{1\over 2})}(y)$  \\ 
s+k ~ even  & $Y_u^k(y) = x P_{s-k-2\over 2}^{(0,{1\over 2})}(y)$  \\ 
\hline 
\end{tabular} 
\end{center}
where $P_n^{(a,b)}$ is Jacobi polynomial.

Due to the relation between Jacobi polynomials and Legendre polynomials $P_n$

\begin{align}
P_n^{(0,-{1\over 2})}(2x^2-1) &= P_{2n}(x) \\
x P_n^{(0,{1\over 2})}(2x^2-1) &= P_{2n+1}(x)
\end{align}
it is of course in agreement with results of the paper \cite{Belavin:2008kv}.

\subsection{Three-point correlation numbers}

Again using (\ref{correlation numbers der}) and (\ref{partition function 2s}) we derive the three-point correlation numbers
\be \label{3-point 2s general}
{\cal Z}_{k_1k_2k_3} = - {Y_u^{k_1}Y_u^{k_2}Y_u^{k_3} \over {d Y_u^0\over dx}  } \Bigg|_{x=1} + \int_0^1 dx Y_u^{k_1k_2} Y_u^{k_3} = - {1\over p} + \int_0^1 dx Y_u^{k_1k_2} Y_u^{k_3}
\ee
where we used the properties of Jacobi polynomials to get: $Y_u^{k}(x=1)=1$, ${d Y_u^0\over dx}\big|_{x=1} = {1\over 2s+1} = {1\over p}$. Besides we assume that $k_1,k_2 \leq k_3$. Two other integral terms in (\ref{3-point 2s general}) with permutations of $k_1, k_2, k_3$ are absent due to (\ref{one point}). 

As we will see the first term reproduces the expression from minimal gravity and the role of the second term is to kill the first term when the fusion rules are violated. 

Also we need not care about the case of odd $k_1+k_2+k_3$. When $k_1+k_2+k_3$ is odd and $< p$ the fusion rules are violated but $Z_{k_1k_2k_3}$ is non-singular (it is an integer positive power of $\mu$). And if $k_1+k_2+k_3$ is odd and $\geq p$ the integral term is automatically zero.

Thus we focus on the case when $k_1+k_2+k_3$ is even. Fusion rules demand 

\be 
\int_0^1 dx Y_u^{k_1k_2} Y_u^{k_3} = \begin{cases} {1\over p} \quad \text{if} \quad k_1+k_2 <k_3 \\ 0 \qquad \text{ if} \quad k_1+k_2 \geq k_3 \end{cases}
\ee 

Again, switching to variables $y$ (\ref{change variable}) we get

\vspace{0.4cm}

\begin{center}
\begin{tabular}{|c|c|}
\hline 
$(s+k_1+k_2)$ ~ odd  & $Y_u^{k_1k_2}(y) = {1\over p}\sum_{n=0}^{s-k_1-k_2-3 \over 2} (4n+1)  P_n^{(0,-{1\over 2})}(y)$  \\
  &  \\
$(s+k_1+k_2)$ ~ even   & $Y_u^{k_1k_2}(y) = {x \over p} \sum_{n=0}^{s-k_1-k_2-4 \over 2} (4n+3) P_n^{(0,{1\over 2})}(y)$  \\
\hline
\end{tabular} 
\end{center}

\vspace{0.5cm}

At this point it is already reasonable to compare some quantities, which are independent on the normalization of fields and normalization of correlators, in the Douglas equation approach with those in Minimal Gravity. Namely, one considers the following combination

\be 
({\cal Z}_{k_1k_2k_3})^2 {\cal Z}_0 \over \prod_{i=1}^3 {\cal Z}_{k_ik_i} 
\ee

When the fusion rules for three-point numbers are satisfied $Z_{k_1k_2k_3} = -{1\over 2s+1}$ and this quantity gives

\be 
\prod_{i=1}^3 (2s-2k_i-1) \over (2s+3)(2s+1)(2s-1)
\ee
which coincides with the value (\ref{3-point MG}) from Minimal Gravity if we take $(q,p) = (2,2s+1), m_i = 1, n_i = k_i+1$.

\subsection{Four-point correlation numbers}

Direct calculation gives

\begin{align}
{\cal Z}_{k_1k_2k_3k_4} &= \left(  - { {d^2 Y_u^0 \over dx^2} \over ({dY_u^0 \over dx})^3}  + {\sum_{i=1}^4 {dY_u^{k_i} \over dx} \over ({dY_u^0 \over dx})^2} -{\sum_{i<j} Y_u^{k_ik_j} \over {dY_u^0 \over dx}} \right) \Bigg|_{x=1}  + \nn \\ &+\int_0^1 dx  (Y_u^{k_1k_2}Y_u^{k_3k_4} + Y_u^{k_1k_3}Y_u^{k_2k_4} +Y_u^{k_1k_4}Y_u^{k_2k_3}) + \nn  \\
&+ \int_0^1 dx (Y_u^{k_1k_2k_3}Y_u^{k_4}+Y_u^{k_1k_2k_4}Y_u^{k_3}+Y_u^{k_1k_3k_4}Y_u^{k_2}+Y_u^{k_2k_3k_4}Y_u^{k_1}) 
\end{align}
We assume that $k_1 \leq k_2 \leq k_3 \leq k_4$. The role of the terms in the third line again is to satisfy the fusion rules by cancelling the terms in the first two lines when the fusion rules are violated. On the other hand the terms in the third line does not appear when the fusion rules are satisfied. Thus to evaluate the four point correlation numbers we need only the terms in the first two lines.

First of all it is convenient to change variables from $x$ to $y$ according to (\ref{change variable})

\begin{align}
{\cal Z}_{k_1k_2k_3k_4} &= \left(  - { 16{d^2 Y_u^0 \over dy^2} + 4 {d Y_u^0 \over dy} \over (4{dY_u^0 \over dy})^3}  + {\sum_{i=1}^4 4{dY_u^{k_i} \over dy} \over (4{ dY_u^0 \over dy})^2} -{\sum_{i<j} Y_u^{k_ik_j} \over 4{dY_u^0 \over dy}} \right) \Bigg|_{y=1}  + \nn \\ &+\int_{-1}^1 {dy \over 2 \sqrt{2} (1+y)^{1\over 2}}  (Y_u^{k_1k_2}Y_u^{k_3k_4} + Y_u^{k_1k_3}Y_u^{k_2k_4} +Y_u^{k_1k_4}Y_u^{k_2k_3})
\end{align}

From the properties of Jacobi polynomials (for both even and odd $s,k,k_1,k_2$)

\begin{align}
(Y_u^0)'(1) &= {2s+1 \over 4}, \qquad \qquad \qquad \quad (Y_u^0)''(1)={(s-1)(s+2)(2s+1) \over 32}, \\
(Y_u^k)'(1) &= {(s-k-1)(s-k) \over 8}, \qquad Y_u^{k_1k_2}(1) = {(s-k_1-k_2-1)(s-k_1-k_2-2) \over 2(2s+1)}
\end{align}
where prime denotes derivative with respect to $y$. Using this we evaluate the four point function in the region where the fusion rules are satisfied

\begin{align}
{\cal Z}_{k_1k_2k_3k_4} &= {1\over 2(2s+1)^2} \Big(  -(s-1)(s+2)-2 + \sum_{i=1}^4 F(k_i + 1)  -  \nn \\ 
&-F(k_{(12|34)})-F(k_{(13|24)})-F(k_{(14|23)}) \Big)
\end{align}
where

\be
F(k) = (s-k-1)(s-k-2), \qquad k_{(ij|lm)} = \min(k_i+k_j,k_l+k_m)
\ee

A reasonable quantity to evaluate here and compare it with the result from the Minimal Gravity is

\be 
({\cal Z}_{k_1k_2k_3k_4}{\cal Z}_0)^2 \over \prod_{i=1}^4 {\cal Z}_{k_ik_i} 
\ee 
and using the properties of Jacobi polynomials and assuming that as in (\ref{4-point condition}) $k_1+k_4 \leq k_2 + k_3$ we get precisely the expression from Minimal Gravity (\ref{4-point MG}) with $(q,p) = (2,2s+1), m_i =1, n_i = k_i+1$.


\section{$\bf (q,p)=(3,3s+p_0)$ Minimal Gravity}

In this case the polynomial $Q$ and the action $S$ are

\be \label{action}
Q = p^3+up+v,\qquad S(u,v) = Res \left( Q^{{p \over 3}+1} + \sum_{n=1}^s \tau_{1,n} Q^{s-n+{p_0 \over 3}} + \sum_{n=s+1}^{p-1} \tau_{1,n} Q^{n-s-{p_0\over 3}} \right)
\ee

One takes then an appropriate solution $(u_*,v_*)$ of the string equations 
\be \label{string equations}
\begin{cases}
 S_u = 0\\
 S_v = 0
\end{cases}
\ee 
where the lowered indices $u,v$ denote the derivatives over $u$ and $v$.
Evaluating the structure constants $C^{\alpha}_{\beta\gamma}$ we find the free energy (\ref{free energy integral}) for this model 

\be \label{free energy}
{\cal F} = {1\over 2} \int_{\gamma(\lambda)} \left( (S_u^2 - {u\over 3} S_v^2)du +2 S_u S_v dv \right)
\ee
where the contour $\gamma(\lambda)$ goes from $(u,v) = (0,0)$ to $(u,v) = (u_*(\lambda), v_*(\lambda))$.
The analysis of the section 3.1 gives
\be 
p \sim \mu^{1\over 6},\quad u\sim \mu^{1\over 3}, \quad v \sim \mu^{1\over 2} \quad Z \sim \mu^{1+{p \over 3}}, \quad S \sim \mu^{ {p +4\over 6}}
\ee
and the dimensions of the times $\tau_{1,k}$
\begin{align} \label{time dim}
\tau_{1,k+1} \sim \mu^{p+3 - |p-3(k+1)| \over 6 } \sim
\begin{cases}
\mu^{k+2 \over 2}, \qquad 0 \leq k \leq s-1 \\
\mu^{s - {k \over 2} + {p_0 \over 3} }, \quad s \leq k \leq p - 2
\end{cases}
\end{align}

\subsection{Resonance relations}
As it was discussed earlier, generally the times introduced in the action are not the times corresponding to the perturbations by primary operators in the Minimal gravity. Namely the resonances are possible. 

By analysing the dimensions of the times (\ref{time dim}) we find possible resonances between the times $\tau_{1,k+1}$ in the Douglas equation approach and coupling constants $\lambda_k$ in minimal gravity up to the second order

\begin{align}
\label{resonance 1}
&\tau_{1,k+1} = \lambda_k + c_k \mu^{k+2 \over 2} + \sum_{\substack{l=1\\(k-l) \in 2Z}}^{s-1} \beta_{kl} \mu^{k-l \over 2} \lambda_l +  \sum C_k^{k_1k_2} \lambda_{k_1} \lambda_{k_2} + \dots , \quad 0 \leq k \leq s-1  \\
&\tau_{1,k+1} = \lambda_k + \sum_{\substack{l=k+2 \\ (l-k) \in 2Z}}^{3s + \alpha -2} \beta_{kl} \mu^{l-k \over 2} \lambda_l +\sum C_k^{k_1k_2} \lambda_{k_1} \lambda_{k_2} +  \dots , \quad s \leq k \leq p - 2 
\label{resonance 2}
\end{align}
Coefficients $C_k^{k_1k_2}$ are non-zero only if the resonance condition is satisfied

\be [\lambda_{k}] = [\lambda_{k_1}] + [\lambda_{k_2}] \ee
for some $k$, where $[\lambda_k]$ is the dimension of the quantity $\lambda_k$. Again by analysing the dimensions (\ref{time dim}) we find that

\be \label{12 rule}
C_k^{k_1k_2} \neq 0, \qquad \text{if} \quad 1\leq k,k_1,k_2 \leq s - 1 \quad \text{or} \quad 
\begin{cases} 0\leq k_1\leq s-1 \\ s \leq k, k_2 \leq p - 2 \end{cases} 
\ee

Once the times $\tau_{1,k+1}$ substituted in (\ref{free energy}) according to (\ref{resonance 1}), (\ref{resonance 2}), one evaluates the n-point correlation number  as

\be \label{n-point def}
Z_{k_1\dots k_n} = {\partial \over \partial \lambda_{k_1}} \dots {\partial \over \partial \lambda_{k_n}} \Bigg|_{\lambda = 0} {\cal F}
\ee

\subsection{Solutions of the string equations}

Now let us describe an appropriate solutions of the string equations (\ref{string equations}). Of course these equations have a number of different solutions. And it is not an easy task to find them all. However we will show that the equations (\ref{string equations}) always possess one particular solution with required properties. We are going to use this solution in the partition function (\ref{free energy}). As a check that we picked out the right solution we will see that it allows to make a correspondence with Minimal Gravity and together with appropriately found resonance relations gives the same results for correlation numbers.
\footnote{We checked in particular examples that other solutions do not allow to make correspondence already at the level of one-point correlation numbers.}

Since the correlation numbers are determined as derivatives of the partition function taken at $\lambda_{m,n}=0$ (except for $\lambda_{1,1}=\mu$), it will be crucial to know the solution $(u_*, v_*)$ of the string equations at such $\lambda_{m,n}$. We will show that one of the equations (\ref{string equations}) is always satisfied by such a $v_*$ that $v_*(\lambda=0)=0$. To demonstrate this we will need the parities of the various parts of the action with respect to the change of the sign of $v$.

Let us substitute the resonance relations (\ref{resonance 1}) and (\ref{resonance 2}) into the action (\ref{action}). We get some expression which can be written as an expansion in $\lambda_k$

\be 
S(u,v) = S^0(u,v) + \sum_{k=1}^{p -2} \lambda_k S^k (u,v) + {1\over 2} \sum_{k_1,k_2=1}^{p -2} \lambda_{k_1}\lambda_{k_2} S^{k_1k_2}(u,v)+\dots
\ee
and dimensional analysis gives

\begin{align}
S^0 &\sim \mu^{p+4 \over 6} 
\\
S^{k_1\dots k_n} &\sim \mu^{{p+4 \over 6} - \sum_{i=1}^{n}{p+3-|p-3(k_i+1)| \over 6}}
\end{align}
Each of the functions $S^{k_1\dots k_n}$ is a polynomial in the variables $u,v,\mu$

\be 
S^{k_1\dots k_n} = \sum_{M,N,K} u^M v^N \mu^K \sim  \mu^{{p+4 \over 6} - \sum_{i=1}^{n}{p+3-|p-3(k_i+1)| \over 6}}
\ee
The dimensions on the right and left hand sides must be the same

\be 
{M\over 3} + {N \over 2} +K = {p+4 \over 6} - \sum_{i=1}^{n}{p+3-|p-3(k_i+1)| \over 6}
\ee
It is equivalent to 

\be 
p+\sum_{i=1}^n k_i -N = 2M+2N+6K + \sum_{i=1}^{n}(p+3-|p-3(k_i+1)|+k_i )
\ee
One can see that the expression on the right hand side is even. Thus $\left(p+\sum_{i=1}^n k_i -N \right)$ is also even. Consequently the functions $S^{k_1\dots k_n}$ have definite parities with respect to $v$

\begin{align} \label{Phi 0 parity}
S^0(u,-v) &= (-1)^{p} S^0(u,v) \\
S^{k_1\dots k_n}(u,-v) &= (-1)^{p+\sum_i k_i} S^{k_1\dots k_n}(u,v) \label{Phi n parity}
\end{align}

Thus, depending on the value of $p$, either $\partial S^0 \over \partial u$ or $\partial S^0 \over \partial v$ is odd function of $v$. Consequently, $v=0$ is always a solution of (\ref{string equations}) at $\lambda =0$. By $u_0$ we denote the solution for $u$ at $\lambda =0$ 

\begin{align}
\begin{cases} S^0_v \big|_{v=0} &\equiv 0 \\ S^0_u \big|_{\substack{v=0 \\ u=u_0}} &= 0 \end{cases} \qquad \text{if} \qquad p \quad \text{is even}
\\
\begin{cases} S^0_u \big|_{v=0} &\equiv 0 \\ S^0_v \big|_{\substack{v=0 \\ u=u_0}} &= 0 \end{cases} \qquad \text{if} \qquad p \quad  \text{is odd}.
\end{align}

Also many of the functions among $S^{k_1 \dots k_n}$ are odd in $v$ and thus vanish at $v=0$. We will extensively use this fact in further considerations.

\subsection{Picking out the contour}

As it was noticed before, the integral (\ref{free energy}) does not depend on the contour of integration. So we can take it at our choice. It is convenient to take the contour $\gamma$ in (\ref{free energy}) as following

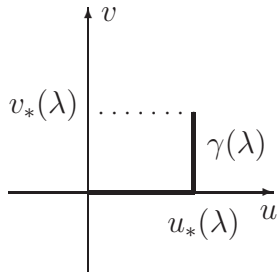
\begin{picture}(120,120)
\put(170,30){\vector(1,0){100}}
\put(200,0){\vector(0,1){100}}
\put(265,20){$u$}
\put(205,95){$v$}
\linethickness{0.5mm}
\put(200,30){\line(1,0){40}}
\put(240,30){\line(0,1){30}}
\put(230,15){$u_*(\lambda)$}
\put(170,60){$v_*(\lambda)$}
\multiput(205,60)(5,0){8}{\circle*{1}}
\put(245,45){$\gamma(\lambda)$}
\end{picture}

This contour has a good property that it reduces the problem to the polynomials of one variable $u$. Indeed, in (\ref{n-point def}) we put $\lambda=0$ after taking derivatives. As we discussed above, we pick out the solution of the string equation such that $(u_*(\lambda=0), v_*(\lambda=0)) = (u_0, 0)$. Thus the contour becomes the line along the $u$-axis

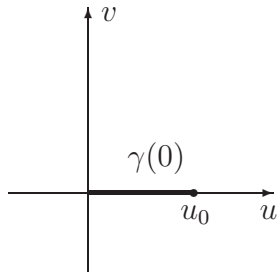
\begin{picture}(120,120)
\put(170,30){\vector(1,0){100}}
\put(200,0){\vector(0,1){100}}
\put(265,20){$u$}
\put(205,95){$v$}
\linethickness{0.5mm}
\put(200,30){\line(1,0){40}}
\put(235,20){$u_0$}
\put(215,40){$\gamma(0)$}
\put(240,30){\circle*{3}}
\end{picture}

So when we evaluate the correlation numbers using (\ref{n-point def}) there will be two kind of terms. The first ones are the non-integral terms, which arise when one of the derivatives act on the upper integration limit in (\ref{free energy}). In these terms after taking $\lambda = 0$ (\ref{n-point def}) the functions must be taken at the point $(u_*(\lambda =0), v_*(\lambda=0)) = (u_0, 0)$. The second ones are the integral terms arising from integrating only the functions under the integral in (\ref{free energy}). After putting $\lambda =0$ the contour of integration becomes $\gamma(0)$ and it is going along the $u$-axis. Thus we must take the functions under the integration at $v=0$. So we see that the correlation numbers are determined by the functions $S^{k_1\dots k_n}$ and their derivatives over $u,v$ taken at $v=0$.

\subsection{Dimensional analysis}

The functions $S^{k_1\dots k_n}(u,0)$ are polynomials in $u$ and $\mu$. It will be important to know their degrees as polynomials in $u$. Dimensional analysis gives

\begin{align}
S^0(u,0) &\sim u^{{p\over 2}+2} +\dots \\
\label{Sk degrees}
S^k(u,0) &\sim \begin{cases} u^{{p\over 2}+2-{3(k+2)\over 2}} +\dots , \qquad 1\leq k \leq s-1 \\
													 u^{-{p\over 2}-1 + {3(k+2) \over 2}}+\dots , \qquad s\leq k \leq p-2 \end{cases} \\
S^{k_1k_2}(u,0) &\sim \begin{cases} u^{{p\over 2}+2-{3(k_1+k_2+4) \over 2}} +\dots , \qquad 1 \leq k_1,k_2 \leq s-1 \\
			    u^{-{p\over 2}-1-{3(k_1-k_2) \over 2}}+\dots, \qquad 
			 	 1\leq k_1 \leq s-1,\quad s \leq k_2 \leq p -2  \end{cases}\\
S^{k_1k_2k_3}(u,0) &\sim \begin{cases}
						u^{{p\over 2}+2-{3(k_1+k_2+k_3+6) \over 2}}+\dots, \qquad 1\leq k_1,k_2,k_3 \leq s-1 \\
						u^{-{p\over 2}-1-{3(k_1+k_2-k_3+2) \over 2}}+\dots, \qquad 1\leq k_1,k_2\leq s-1, \quad s\leq k_3\leq p-2 \\
						u^{-{3p\over 2}-4-{3(k_1-k_2-k_3-2) \over 2}}+\dots, \qquad 1\leq k_1\leq s-1, \quad s\leq k_2,k_3\leq p-2 \\
						u^{-{5p\over 2}-7+{3(k_1+k_2+k_3+6) \over 2}}+\dots, \qquad s\leq k_1,k_2,k_3\leq p-2 
													\end{cases}
\label{Skkk degrees}
\end{align}
\begin{align}
S^{k_1k_2k_3k_4}(u,0) &\sim \begin{cases} 
						u^{{p\over 2}+2-{3(k_1+k_2+k_3+k_4+8)\over 2}}+\dots, \qquad 1\leq k_1,k_2,k_3,k_4\leq s-1 \\ 
						u^{-{p\over 2}-1-{3(k_1+k_2+k_3-k_4+4)\over 2}}+\dots, \qquad 1\leq k_1,k_2,k_3\leq s-1, \quad s\leq k_4\leq p-2 \\ 
						u^{-{3p\over 2}-4-{3(k_1+k_2-k_3-k_4)\over 2}}+\dots, \qquad 1\leq k_1,k_2\leq s-1, \quad s\leq k_3,k_4\leq p-2 \\
						u^{-{5p\over 2}-7-{3(k_1-k_2-k_3-k_4-4)\over 2}}+\dots, \qquad 1\leq k_1\leq s-1,\quad s\leq k_2,k_3,k_4\leq p-2 \\
						u^{-{7p\over 2}-10+{3(k_1+k_2+k_3+k_4+8)\over 2}}+\dots, \qquad s\leq k_1,k_2,k_3,k_4\leq p-2   \end{cases} 		
\end{align}
where the dots represent the terms involving cosmological constant $\mu$. Since $\mu \sim u^3$ the degrees of $u$ in each polynomial go with the step $3$. Also it is worth noticing that if the degree of the polynomial $S^{k_1\dots k_n}$ is non-integer or negative, it means that it is actually zero. This can be seen from the expression for the action (\ref{action}). Indeed, it would just mean that the corresponding residue vanishes.

Also we disregard all the correlation numbers which are proportional to the integer powers of the cosmological constant $\mu$ as non-universal. So it is important to know the dimensions of the correlation numbers in order to check whether we have a non-trivial fusion rule for them

\begin{align}
\l O_k \r &\sim \begin{cases} \mu^{{p\over 3} - {k \over 2}}, \quad 1\leq k \leq s-1 
															\\ \mu^{k+2 \over 2}, \quad s \leq k \leq p-2 \end{cases} \\
\l O_{k_1}O_{k_2} \r &\sim \begin{cases} \mu^{{p\over 3}+1- {k_1+k_2+4 \over 2}}, \quad 1\leq k_1, k_2 \leq s-1 
															\\ \mu^{k_2-k_1 \over 2}, \quad 1\leq k_1 \leq s-1, \quad  s \leq k_2 \leq p- 2 \\
															\mu^{-{p\over 3}-1+{k_1+k_2+4 \over 2} }, \quad s \leq k_1,k_2 \leq p-2 \end{cases} \\
\label{dimensions 3-point}
\l O_{k_1}O_{k_2}O_{k_3} \r &\sim \begin{cases} \mu^{{p \over 3}+1 - {k_1+k_2+k_3+6 \over 2}}, \quad 1\leq k_1, k_2,k_3 \leq s-1 
															\\ \mu^{k_3-k_1-k_2-2 \over 2}, \quad 1\leq k_1,k_2 \leq s-1, \quad  s \leq k_3 \leq p- 2 \\
												\mu^{-{p\over 3}-1+{k_2+k_3-k_1+2 \over 2} }, \quad 1\leq k_1\leq s-1,\quad s \leq k_2,k_3 \leq p-2 \\
												\mu^{-{2p\over 3}-2+{k_1+k_2+k_3+6 \over 2} }, \quad  s \leq k_1,k_2,k_3 \leq p- 2 \\
													 \end{cases} \\
\l O_{k_1}O_{k_2}O_{k_3}O_{k_4} \r &\sim \begin{cases} 
\mu^{{p\over 3}+1-{k_1+k_2+k_3+k_4+8 \over 2}}, \quad 1\leq k_1,k_2,k_3,k_4 \leq s-1, \\
\mu^{k_4 - k_1 - k_2 - k_3 -4 \over 2} ,\quad 1\leq k_1,k_2,k_3 \leq s-1,\quad s\leq k_4 \leq p-2, \\
\mu^{-{p\over 3}-1+{k_3 + k_4 -k_1-k_2 \over 2}}, \quad 1\leq k_1,k_2 \leq s-1,\quad s\leq k_3,k_4\leq p-2, \\
\mu^{-{2p\over 3}-2 +{k_2+k_3 + k_4 -k_1+4 \over 2}}, \quad 1\leq k_1\leq s_1, \quad s\leq k_2,k_3,k_4 \leq p-2, \\ 
\mu^{-p-3+{k_1+k_2+k_3+k_4+8 \over 2}}, \quad s\leq k_1,k_2,k_3,k_4 \leq p-2
\end{cases} \label{dimension 4-point}
\end{align}

Now we are going to evaluate zero-, one-, two, three- and four-point correlation numbers.

\subsection{Zero-, one- and two-point numbers}

Before going into details let us point out a few issues.

First of all note that, when evaluating one- and two-point numbers,  we do not need to differentiate over the limits of integration since at $(u_*,v_*)|_{\lambda=0}$ these terms give zero because of the string equation.

Next, when taking $\lambda =0$ after differentiating over $\lambda$'s in (\ref{n-point def}), one finds that many of the functions $S_u^{k_1\dots k_n}$, $S_v^{k_1\dots k_n}$ are odd in $v$ due to (\ref{Phi 0 parity}) - (\ref{Phi n parity}) and thus vanish at $v=0$ .

Taking these things into account one evaluates zero-, one- and two-point correlation numbers. The result is summarized in the table

\vspace{0.5cm}
\begin{center}
\begin{tabular}{|c|c|c|} 
\hline 
  & $p$ ~ even & $p$ - odd \\ 
\hline
 & & \\
 & $Z_0 = {1\over 2}\int_0^{u_0}(S_u^0)^2du$  & $Z_0=-{1\over 6}\int_0^{u_0}  (S_v^0)^2 u du $ \\ 
 & & \\
 & $Z_k = \int_0^{u_0}S_u^0 S_u^k du$ & $Z_k = - {1\over 3}\int_0^{u_0}  S_v^0 S_v^k u du$ \\
 & & \\ 
$k_1,k_2$ ~ even & $Z_{k_1k_2} = \int_0^{u_0}(S_u^{k_1} S_u^{k_2} + S_u^0 S_u^{k_1k_2}  )du$ & $Z_{k_1k_2} = -{1\over 3}\int_0^{u_0}(S_v^{k_1} S_v^{k_2} + S_v^0 S_v^{k_1k_2}  )udu$ \\ 
 & & \\
 $k_1,k_2$ ~ odd & $Z_{k_1k_2} = \int_0^{u_0}(S_u^0 S_u^{k_1k_2} - {u\over 3}S_v^{k_1} S_v^{k_2}  )du$ & $Z_{k_1k_2} = \int_0^{u_0}(S_u^{k_1} S_u^{k_2} - {u\over 3}S_v^0 S_v^{k_1k_2}  )du$ \\ 

 & & \\
$k_1+k_2$ ~ odd & $Z_{k_1k_2}=0$ & $Z_{k_1k_2}=0$ \\ 
\hline 
\end{tabular}   
\end{center}
where all the polynomials are taken at $v=0$ since the contour of integration goes along the $u$-axis.

Now for both $p$ - even or $p$ - odd the analysis is similar to the case of Lee-Yang series $(2,2s+1)$. 
Similarly to that case we switch from $S(u,0)$ and $u$ to dimensionless quantities $Y(x)$ and $x={u\over u_0}$ respectively. We do not need the precise relation between dimensional action $S$ and dimensionless $Q$ and do not write it down in order not to confuse the reader with unnecessary messy coefficients. Also we implicitly switched from $\lambda_k$ to dimensionless quantities $s_k$ and we now assume that $\partial_k = {\partial \over \partial s_k}$. 

Also as in the case of Lee-Yang series we use the variable $y$

\be 
{y+1 \over 2} = x^3, \qquad dx= {dy \over 3\sqrt[3]{2} (1+y)^{2\over 3}}
\ee

Following the logic similar to Lee-Yang series we receive

\vspace{0.5cm}
\begin{center}
\begin{tabular}{|c|c|}
\hline 
 $p$ ~ even & $Y_u^0 = x^{2(p_0-1)} ( P_{s-p_0 +2 \over 2}^{(0,{2\over 3}(2p_0-3))}(y) - P_{s-p_0 \over 2}^{(0,{2\over 3}(2p_0-3))}(y))$  \\ & \\
  $p$ ~ odd & $Y_v^0 = x^{2-p_0}(P_{s+p_0 -1 \over 2}^{(0,{2\over 3}(1-p_0))}(y) - P_{s+p_0 -3 \over 2}^{(0,{2\over 3}(1-p_0))}(y)) $ 
\\  & \\
$(p+k)$ ~ even, $k<s$ & $Y_u^k =x^{2(p_0-1)} P_{s-k-p_0 \over 2}^{(0,{2\over 3}(2p_0-3))}(y)$   \\  &  \\
$(p+k)$ ~ odd, $k<s$ & $Y_v^k = x^{2-p_0} P_{s-k+p_0 -3 \over 2}^{(0,{2\over 3}(1-p_0))}(y) $  \\  & \\
$(p+k)$ ~ even, $k\geq s$ & $Y_u^k =x^{2(2-p_0)} P_{k-s+p_0 -2 \over 2}^{(0,{2\over 3}(3-2p_0))}(y)$  \\  & \\
$(p+k)$ ~ odd, $k\geq s$ & $Y_v^k = x^{p_0 -1 } P_{k-s-p_0 +1 \over 2}^{(0,{2\over 3}(p_0-2))}(y) $  \\
\hline
\end{tabular} 
\end{center}
where all the polynomials are evaluated at $v=0$.

\subsection{Three-point correlation numbers}

For the three-point correlation numbers a direct calculation gives  

\begin{align}
Z_{k_1k_2k_3} =  (S_u^{k_1}&S_u^{k_2} - {u_0 \over 3} S_v^{k_1}S_v^{k_2}) \partial_{k_3}u_* + S_u^{k_1}S_v^{k_2} \partial_{k_3}v_* + \nn \\ &+ \int_0^{u_0} du (S_u^{k_1}S_u^{k_2k_3} - {u\over 3} S_v^{k_1}S_v^{k_2k_3}) + \int_0^{u_0} du( S_u^0 S_u^{k_1k_2k_3} -{u\over 3} S_v^0 S_v^{k_1k_2k_3}) + permutations
\end{align}
Taking derivative of the equations (\ref{string equations}) one gets at $s_k=0$ 

\begin{align}
S_u^k + S_{uu}^0 \partial_k u_* + S_{uv}^0 \partial_k v_* &= 0 \\
S_v^k + S_{vu}^0 \partial_k u_* + S_{vv}^0 \partial_k v_* &= 0 \\
\end{align}
So using the parities of polynomials $S^{k_1\dots k_n}$ one has at $v=0$

\begin{center}
\begin{tabular}{|c|c|c|}
\hline 
 & $p$ ~ even & $p$ ~ odd \\ \hline && \\

$k$ ~ even & $\partial_k u_* = - {S_u^{k} \over S_{uu}^0}$ & $\partial_k u_* = - {S_v^{k} \over S_{uv}^0}$ \\ && \\
$k$ ~ odd & $\partial_k v_* = - {S_v^{k} \over S_{vv}^0}$ & $\partial_k v_* = - {S_u^{k} \over S_{uv}^0}$ \\ 
\hline 
\end{tabular} 
\end{center}
where for each case only non-zero derivative is represented. For instance for even $p$ and $k$ we have $\p_k v_* = 0$.

Now using this expressions for derivatives of $(u_*,v_*)$ in three-point numbers and, again, using the parities of $S^{k_1\dots k_n}$ one gets the following result for $Z_{k_1k_2k_3}$ at $v=0$

\begin{center}
\begin{tabular}{|c|c|c|}
\hline 
 & $p$ ~ even & $p$ ~ odd \\ \hline && \\
$k_1,k_2,k_3$ ~ even  & $- {S_u^{k_1}S_u^{k_2}S_u^{k_3} \over S_{uu}^{0}} + \int  S_u^{0}S_u^{k_1k_2k_3}$ & ${u_0 \over 3} {S_v^{k_1}S_v^{k_2}S_v^{k_3} \over S_{uv}^{0}}- \int {u\over 3} S_v^{0}S_v^{k_1k_2k_3} - $ \\ 
  & $+\int (S_u^{k_1}S_u^{k_2k_3}+{\rm permutations})$ & $-\int {u\over 3} (S_v^{k_1}S_v^{k_2k_3}+{\rm permutations})$ \\ \hline 
$k_1,k_2$ ~ even && \\
$k_3$ ~ odd & 0 & 0 \\ \hline && \\
$k_1$ ~ even & $ {u_0 \over 3} {S_u^{k_1}S_v^{k_2}S_v^{k_3} \over S_{uu}^{0}} - {S_u^{k_1}S_v^{k_2}S_v^{k_3} \over S_{vv}^{0}} + \int S_u^{0}S_u^{k_1k_2k_3} +  $ & $- {S_v^{k_1}S_u^{k_2}S_u^{k_3} \over S_{uv}^{0}} - \int {u\over 3} S_v^{0}S_v^{k_1k_2k_3} + $ \\ 
$k_2,k_3$ ~ odd  & $+\int (S_u^{k_1}S_u^{k_2k_3}-{u\over 3}(S_v^{k_2}S_v^{k_1k_3}+S_v^{k_3}S_v^{k_1k_2}))$ & $+\int (S_u^{k_2}S_u^{k_1k_3}+S_u^{k_3}S_u^{k_1k_2}-{u\over 3}S_v^{k_1}S_v^{k_2k_3})$ \\ \hline && \\
$k_1,k_2,k_3$ ~ odd & 0 & 0 \\ \hline
\end{tabular} 
\end{center}
where the off-integral terms are evaluated at $u=u_0$ and all the integrals are taken from $0$ to $u_0$.

The simplest case to analyse here is when $p$ is even and all $k$ are even, that is, in the upper left column of the tableau. So let us concentrate on this case

\be \label{three-point 3s}
Z_{k_1k_2k_3} = - {S_u^{k_1}S_u^{k_2}S_u^{k_3} \over S_{uu}^{0}}\Big|_{u=u_0} + \int_0^{u_0}  S_u^{0}S_u^{k_1k_2k_3}du+\int_0^{u_0} (S_u^{k_1}S_u^{k_2k_3}+S_u^{k_2}S_u^{k_1k_3} + S_u^{k_3}S_u^{k_1k_2})du.
\ee 

$\bullet$  $1\leq k_1,k_2,k_3\leq s-1$; $p,k$~-- even

In this case the 3-point correlation numbers are singular due to eq. (\ref{dimensions 3-point}). The expression for them reduces to (we assume that $k_3>k_1,k_2$) 
\be 
{\cal Z}_{k_1k_2k_3} = - {Y_u^{k_1}Y_u^{k_2}Y_u^{k_3} \over (Y_{u}^{0})'}\Big|_{x=1} +\int_0^{1} Y_u^{k_3}Y_u^{k_1k_2}dx
\ee
where we switched to dimensionless quantities and prime denotes the derivative over $x$. Two other integral terms are absent because the degree of the polynomial $S^{k_2k_3}_u$ is less than that of the Jacobi polynomial $S^k_u$. The other term is treated similarly. Thus to satisfy fusion rules (\ref{conf rule 3}) one needs the following condition to hold

\be 
\int_0^{1} Y_u^{k_3}Y_u^{k_1k_2}dx = \begin{cases} 0, \qquad\qquad\qquad\qquad\quad \text{ if} \quad k_3\leq k_1+k_2 \\ {Y_u^{k_1}Y_u^{k_2}Y_u^{k_3} \over (Y_{u}^{0})'}\Big|_{x=1} = {1\over p}, \qquad \text{if} \quad k_3 > k_1+k_2 \end{cases}
\ee 

This determines 

\be \label{Qk1k2 1} \boxed{ 
Y^{k_1k_2}_u = {1\over p}\sum_{k=0}^{s-k_1-k_2-p_0 -2 \over 2}(6k+4p_0 - 3)x^{2(p_0 -1)}P_k^{(0,{2\over 3}(2p_0 - 3))}, \qquad 1\leq k_1,k_2\leq s-1}
\ee

Now we can evaluate the correlation numbers when the fusion rules are satisfied

\begin{align} \label{zero-point corr numbe}
Z_0 &= {p \over (p+3)(p-3)} \\
Z_{kk} &= \begin{cases} {1\over p - 3(k+1)}, \qquad 1\leq k \leq s-1, \quad k ~ \text{even} \\ {1\over 3(k+1)- p}, \qquad s\leq k \leq p-2, \quad k~ \text{even}  \end{cases} \\
Z_{k_1k_2k_3} &= -{1\over p}, \quad\qquad\qquad 1\leq k_i\leq p-2, \quad k ~ \text{even}.
\label{3-point value}
\end{align}

The quantity that doesn't depend on the normalization of the operators and correlators is 

\be \label{normalized three-point}
{(Z_{k_1k_2k_3})^2 Z_0 \over \prod_{i=1}^3 Z_{k_ik_i}} = {\prod_{i=1}^3 |p -  k_i q| \over p(p+q)(p-q)}
\ee 
where $p=3s+p_0, q=3$. It is in agreement with the direct calculation of the integrals over the moduli space in Minimal Gravity (\ref{3-point MG}).

For the other ranges of parameters $k_1,k_2,k_3$ we will see that the three-point correlation numbers are always given by (\ref{3-point value}) when the fusion rules are satisfied. Thus, the result for those cases also coincides with the direct calculations in Liouville Minimal Gravity.

$\bullet$ {$1\leq k_1,k_2\leq s-1$, $s\leq k_3 \leq p-2$; $p,k_i$~-- even}

In this case $k_1+k_2 \leq 2(s-2) \leq p-2$. Thus the fusion rules (\ref{conf rule 3}) are violated if $k_3>k_1+k_2$, or since all $k_i$ are even, it is equivalent to $k_3 \geq k_1+k_2+2$. Thus the 3-point correlation numbers are non-singular (\ref{dimensions 3-point}) when the fusion rules are violated.

On the other hand if the fusion rules are satisfied then the 3-point correlation numbers are singular and the expression for them reduces to

\be
{\cal Z}_{k_1k_2k_3} = - {Y_u^{k_1}Y_u^{k_2}Y_u^{k_3} \over (Y_{u}^{0})'}\Big|_{x=1} + \int_0^1  Y_u^{0}Y_u^{k_1k_2k_3}dx+\int_0^1 (Y_u^{k_1}Y_u^{k_2k_3}+Y_u^{k_2}Y_u^{k_1k_3} + Y_u^{k_3}Y_u^{k_1k_2})dx.
\ee

The integral term here is actually zero, when the fusion rules are satisfied, i.e. when $k_3 \leq k_1+k_2$. Indeed, consider the term $\int_0^1 Y_u^{0}Y_u^{k_1k_2k_3}dx$. When $k_3 \leq k_1+k_2$ the degree (\ref{Skkk degrees}) of the polynomial $Y_u^{k_1k_2k_3}$ is less than zero, ${3(k_3-k_1-k_2-s-3)\over 2}+{-1-p_0 \over 2} \leq -{3(s+3)\over 2}+{-1-p_0 \over 2} <0$. This actually means that the polynomial vanishes in this case because the functions $Y^{k_1\dots k_n}(x)$ were defined as coefficients of the expansion in $\lambda_k$ of the polynomial $Y(x)$. So the functions $Y^{k_1\dots k_n}(x)$ must be polynomials. Similarly, when $k_3\leq k_1+k_2$, the polynomials $Y_u^{k_1k_2}, Y_u^{k_2k_3}, Y_u^{k_1k_3}$ vanish. Thus we arrive at the following expression for the three-point correlation numbers when the fusion rules are satisfied

\be 
{\cal Z}_{k_1k_2k_3} = - {Y_u^{k_1}Y_u^{k_2}Y_u^{k_3} \over (Y_{u}^{0})'}\Big|_{x=1} = -{1\over p}
\ee

$\bullet$ {$1\leq k_1\leq s-1$, $s\leq k_2,k_3 \leq p-2$; $p,k_i$~-- even}

In this case the 3-point numbers are once again singular (\ref{dimensions 3-point}). The expression for them is

\be \label{three-point 1 2}
{\cal Z}_{k_1k_2k_3} = - {Y_u^{k_1}Y_u^{k_2}Y_u^{k_3} \over (Y_{u}^{0})'}\Big|_{x=1} +\int_0^{1} Y_u^{k_2}Y_u^{k_1k_3}dx.
\ee
If additionally $k_1+k_2 \leq p-2$ then the fusion rules (\ref{conf rule 3}) require that the polynomials $Y_u^{k_1k_3}$ satisfy 

\be \label{fusion equation 3-point 1}
\int_0^{1} Y_u^{k_2}Y_u^{k_1k_3}dx = \begin{cases} 0, \qquad \text{if} \quad k_3\leq k_1+k_2 \\ {1\over p}, \qquad \text{if} \quad k_3 > k_1+k_2 \end{cases}, \quad k_1+k_2 \leq p-2.
\ee 
If $k_1+k_2 > p-2$ then

\be \label{fusion equation 3-point 2}
\int_0^{1} Y_u^{k_2}Y_u^{k_1k_3}dx = \begin{cases} 0, \qquad \text{if} \quad k_1+k_2+k_3\leq 2(p-2) \\ {1\over p}, \quad \text{if} \quad k_1+k_2+k_3 > 2(p-2) \end{cases}, \quad k_1+k_2 >p-2.
\ee 
However, notice that, when $k_3 \leq k_1+k_2$ the integral $\int_0^{1} Y_u^{k_2}Y_u^{k_1k_3}dx$ automatically vanishes since the degree of the polynomial $Y_u^{k_1k_3}$ is less then the degree of the orthogonal Jacobi polynomial $Y_u^{k_2}$. Thus we can't satisfy the condition (\ref{fusion equation 3-point 2}), since $k_1+k_2 > p-2 \geq k_3$. This is the first time where we encounter problems with satisfying the selection rules. Nevertheless, despite the fact that we can not satisfy all of the selection rules, we will see that after satisfying a part of the selection rules we can get some sensible results. For the three-point correlation numbers this will allow us to find all the polynomials $Y_u^{k_1k_2}$. As a check that we get the right expressions for these polynomials we will calculate the four-point correlation numbers which satisfy the fusion rules (and thus are not zero) and involve the polynomials $Y_u^{k_1k_2}$. We will see that they coincide with four-point correlation numbers from Minimal Liouville Gravity (\ref{4-point MG}).

Thus the only non-trivial condition which we have for the polynomials $Y_u^{k_1k_3}$ is (\ref{fusion equation 3-point 1}).
This determines the polynomials $Y_u^{k_1k_3}$.

\be \label{Qk1k2 2} \boxed{
Y^{k_1k_3}_u = {1\over p}\sum_{k=0}^{k_3-k_1-s+p_0 -4 \over 2}(6k+9-4p_0 )x^{2(2-p_0)}P_k^{(0,{2\over 3}(3-2p_0))}, \qquad 
\begin{cases} 
1\leq k_1\leq s-1, \\ s\leq k_3 \leq p-2
\end{cases}
}
\ee

$\bullet$ {$s\leq k_1,k_2,k_3 \leq p-2$; $p,k_i$~-- even}

In this case the three-point correlation numbers are singular (\ref{dimensions 3-point}) and the expression for them reduces to

\be 
Z_{k_1k_2k_3} = - {Y_u^{k_1}Y_u^{k_2}Y_u^{k_3} \over (Y_{u}^{0})'}\Big|_{x=1} + \int_0^1  Y_u^{0}Y_u^{k_1k_2k_3}dx.
\ee
Thus to satisfy the conformal and selection rules we need

\be \label{three point last 1}
\int_0^1  Y_u^{0}Y_u^{k_1k_2k_3}dx = \begin{cases}
0, \quad \text{if} \quad k_3 \leq k_1+k_2 \\
{1\over p}, \quad \text{if} \quad k_3> k_1+k_2
\end{cases}, \quad {\rm for}\quad k_1+k_2 \leq p-2
\ee
and

\be \label{three point last 2}
\int_0^1  Y_u^{0}Y_u^{k_1k_2k_3}dx = \begin{cases}
0, \quad \text{if} \quad k_1+k_2+k_3 \leq 2(p-2) \\
{1\over p}, \quad \text{if} \quad k_1+k_2+k_3> 2(p-2)
\end{cases}, \quad {\rm for}\quad k_1+k_2 > p-2.
\ee

Observe that the integral $\int_0^1  Y_u^{0}Y_u^{k_1k_2k_3}dx$ vanishes automatically when $k_1+k_2+k_3 \leq 2(p-2)$ for the same reason as did the integral $\int_0^1  Y_u^{k_2}Y_u^{k_1k_3}dx$ in (\ref{fusion equation 3-point 2}). Thus we see that if $k_1+k_2 \leq p-2$ then $k_1+k_2+k_3 \leq 2(p-2)$ and the condition (\ref{three point last 1}) cannot be satisfied because the integral $\int_0^1  Y_u^{0}Y_u^{k_1k_2k_3}dx$ vanishes automatically.

This finishes the analysis of the three-point correlation numbers and we shall move on to the four-point case now.

\subsection{Four-point correlation numbers}

Direct calculation gives for even $p,k_i$
\be \label{4-point 3s}
Z_{k_1k_2k_3k_4} = Z_{k_1k_2k_3k_4}^{(0)}+Z_{k_1k_2k_3k_4}^{(I)}
\ee
where
\begin{align} \nn
Z_{k_1k_2k_3k_4}^{(0)} =  & \left( - {(Y_u^0)'' \over ((Y_u^0)')^3}  + {\sum_{i=1}^4 (Y_u^{k_i})' \over ((Y_u^0)')^2} -{\sum_{i<j} Y_u^{k_ik_j} \over (Y_u^0)'} \right)\Bigg|_{x=1} + \\ &+\int_0^1 dx  (Y_u^{k_1k_2}Y_u^{k_3k_4} + Y_u^{k_1k_3}Y_u^{k_2k_4} +Y_u^{k_1k_4}Y_u^{k_2k_3})  \\
Z_{k_1k_2k_3k_4}^{(I)} = & \int_0^1 dx (Y^{k_1k_2k_3}_u Y^{k_4}_u + Y^{k_1k_2k_4}_u Y^{k_3}_u + Y^{k_1k_3k_4}_u Y^{k_2}_u + Y^{k_2k_3k_4}_u Y^{k_1}_u + Y^{0}_u Y^{k_1k_2k_3k_4}_u )
\end{align}
where prime denotes the $x$-derivative. We assume that $k_1\leq k_2\leq k_3\leq k_4$. 

$\bullet$ {$1\leq k_1,k_2,k_3,k_4 \leq s-1$; $p,k_i$ -- even}

First of all the dimensional analysis gives that in this case the 4-point correlation numbers are singular (\ref{dimension 4-point}). Polynomials involved in $Z_{k_1k_2k_3k_4}^{(0)}$ are known from the previous consideration and polynomials involved in $Z_{k_1k_2k_3k_4}^{(I)}$ are not known yet. 

For $k_1\leq k_2\leq k_3\leq k_4$ the orthogonality of Jacobi polynomials gives (the other terms are zero because they involve the integral of the product of the Jacobi polynomial and the polynomial whose degree is less then that of the Jacobi polynomial)
\be
Z_{k_1k_2k_3k_4}^{(I)} = \int_0^1 dx Y^{k_1k_2k_3}_u Y^{k_4}_u 
\ee

Since $Y^{k_4}_u$ is the Jacobi polynomial, $Z_{k_1k_2k_3k_4}^{(I)}$ automatically equals zero when the degree of the polynomial $Y^{k_1k_2k_3}_u$ is less then that of $Y^{k_4}_u$ or, equivalently, when (see (\ref{Sk degrees}), (\ref{Skkk degrees})) $k_4 \leq k_1+k_2+k_3+2$. 

The fusion rules (\ref{conf rule 4}) in this case are 

\be 
Z_{k_1k_2k_3k_4}  =  0, \qquad \text{iff} \quad k_4 > k_1+k_2+k_3 
\ee

Thus to satisfy the fusion rules we need 

\be \label{fusion444}
\int_0^1 dx Y^{k_1k_2k_3}_u Y^{k_4}_u = \begin{cases} 0, \qquad k_4 \leq k_1+k_2+k_3  \\ -Z_{k_1k_2k_3k_4}^{(0)}, \qquad k_4> k_1+k_2+k_3 \end{cases}
\ee

Using the properties of Jacobi polynomials (see Appendix B) to evaluate $Z_{k_1k_2k_3k_4}^{(0)}$ we find that 


\begin{align} \label{4-point non-normalized 1}
Z_{k_1k_2k_3k_4}^{(0)} = \begin{cases} {1\over 4p^2} (2+k_1+k_2+k_3-k_4)(2p-3\sum_{i=1}^4(k_i+1)), \qquad \text{if} \quad k_4>k_1+k_2+k_3 \\ 
									{1\over 2p^2} (1+k_1)(2p-3\sum_{i=1}^4(k_i+1)), \qquad \text{if} \quad k_1+k_4 \leq k_2+k_3 \end{cases} 
\end{align}

Thus we find from the fusion rule (\ref{fusion444}) 

\begin{align} \label{Qk1k2k3 1} \boxed{
Y_u^{k_1k_2k_3} = {1\over p^2}\sum_{k=0}^{s-k_1-k_2-k_3-p_0 -4 \over 2}c_k x^{2(p_0-1)} P_k^{(0,{2\over 3}(2p_0-3))}(y), \qquad
1\leq k_1,k_2,k_3 \leq s-1
}
\end{align}
where
\be
c_k = {1\over 4}(6k+4p_0-3)(2k+\sum_{i=1}^3 k_i -s+p_0+2)(2p-3\sum_{i=1}^3 k_i + 3(2k-s+p_0)-12)
\ee

Also, when $k_1+k_4 \leq k_2+k_3$ and the fusion rules are satisfied we find from (\ref{4-point non-normalized 1}) that

\be \label{4-point normalized 1}
{Z_{k_1k_2k_3k_4}Z_0 \over \left( \prod_{i=1}^4 Z_{k_ik_i} \right)^{1\over 2} } = {\prod_{i=1}^4 |p-3(k_i+1)|^{1\over 2}\over 2p(p+3)(p-3)} (1+k_1)(2p-3\sum_{i=1}^4(k_i+1))
\ee
exactly coincides with the result of direct calculation \cite{Belavin:2006ex} in Minimal Gravity (\ref{4-point MG}), where $m_i=1, n_i=k_i+1, q=3$.

So far the calculations of four-point correlation numbers are in full agreement with the Liouville Minimal Gravity. However, in the next cases we will have only partial agreement.

$\bullet$ {$1\leq k_1,k_2,k_3\leq s-1, s\leq k_4 \leq p-2$; $p,k_i$ -- even}

The dimensional analysis (\ref{dimension 4-point}) gives that in this case if $k_4 < k_1+k_2+k_3+4$ then the 4-point numbers are singular. Meanwhile the selection rules (\ref{conf rule 4}) are violated if $k_4 > k_1+k_2+k_3$. Thus there is a ``window" at $k_4 = k_1+k_2+k_3+2$ when it needs to be checked that the 4-point numbers are zero. So one needs to verify that $Z^{(0)}_{k_1k_2k_3k_4}+Z^{(I)}_{k_1k_2k_3k_4}=0$ when $k_4 =k_1+k_2+k_3+2$. Besides, when $k_4=k_1+k_2+k_3+2$ there are no counter terms $Z^{(I)}_{k_1k_2k_3k_4}$ involving polynomials $Y_u^{k_1k_2k_3}$. The expression for the four-point correlation numbers reduce to

\begin{align} \label{4-point 2 case}
Z_{k_1k_2k_3k_4} =  & \left( - {(Y_u^0)'' \over ((Y_u^0)')^3}  + {\sum_{i=1}^4 (Y_u^{k_i})' \over ((Y_u^0)')^2} -{\sum_{i<j} Y_u^{k_ik_j} \over (Y_u^0)'} \right)\Bigg|_{x=1} + \\ \nn &+\int_0^1 dx  (Y_u^{k_1k_2}Y_u^{k_3k_4} + Y_u^{k_1k_3}Y_u^{k_2k_4} +Y_u^{k_1k_4}Y_u^{k_2k_3}) 
\end{align}
All the polynomials in this expression are known from the previous discussion. Let us compare this expression with the result (\ref{4-point MG}) of the direct calculation in Minimal Gravity. It is assumed that the formula (\ref{4-point MG}) is correct under the assumption (\ref{4-point condition}) $k_1+k_4 \leq k_2+k_3$. One can check that under this condition the degrees of the polynomials $Y^{k_2k_3}_u, Y^{k_2k_4}_u, Y^{k_3k_4}$ (see (\ref{Qk1k2 1}), (\ref{Qk1k2 2})) are less than zero. This just means that these polynomials equal to zero when $k_1+k_4 \leq k_2+k_3$. Thus the expression (\ref{4-point 2 case}) for the 4-point correlation numbers reduces to 
\begin{align}
Z_{k_1k_2k_3k_4} =  & \left( - {(Y_u^0)'' \over ((Y_u^0)')^3}  + {\sum_{i=1}^4 (Y_u^{k_i})' \over ((Y_u^0)')^2} -{\sum_{i=2}^4 Y_u^{k_1k_i} \over (Y_u^0)'} \right)\Bigg|_{x=1}
\end{align}

Using explicit expressions for the polynomials we find that 

\be 
Z_{k_1k_2k_3k_4} = -{3\over 2} (1+k_1)(2+k_1+k_2+k_3-k_4).
\ee
This gives for the normalized 4-point function 

\be \label{4-point normalized 5}
{Z_{k_1k_2k_3k_4}Z_0 \over \left( \prod_{i=1}^4 Z_{k_ik_i} \right)^{1\over 2} } = {\prod_{i=1}^4 |p-3(k_i+1)|^{1\over 2}\over 2p(p+3)(p-3)} 3(1+k_1)(k_4-k_1-k_2-k_3-2)
\ee
which exactly coincides with (\ref{4-point MG}).

$\bullet$ {$1\leq k_1,k_2\leq s-1, s\leq k_3,k_4 \leq p-2$; $p,k_i$ -- even}

Dimensional analysis (\ref{dimension 4-point}) gives that the 4-point correlation numbers are singular.  

The expression (\ref{4-point 3s}) reduces to 

\be 
Z_{k_1k_2k_3k_4} = Z^{(0)}_{k_1k_2k_3k_4}+ \int_0^1 dx Y_u^{k_1k_2k_4}Y_u^{k_3}
\ee

Thus to satisfy the fusion rules (\ref{conf rule 4}) we need
\be \label{fusion 4-point 3s}
\int_0^1 dx Y_u^{k_1k_2k_4}Y_u^{k_3} = \begin{cases}
 0, \qquad k_4 \leq k_1+k_2+k_3 \\ 
 -Z^{(0)}_{k_1k_2k_3k_4}, \qquad  k_4 > k_1+k_2+k_3 
\end{cases}, \quad \text{when}\quad k_1+k_2+k_3 \leq p-2
\ee
and

\be \label{fusion 4-point 3s 2}
\int_0^1 dx Y_u^{k_1k_2k_4}Y_u^{k_3} = \begin{cases}
 0, \qquad k_1+k_2+k_3+k_4 \leq 2(p-2) \\ 
 -Z^{(0)}_{k_1k_2k_3k_4}, \quad k_1+k_2+k_3+k_4 > 2(p-2) 
\end{cases}, \quad k_1+k_2+k_3 > p-2.
\ee
Here, similarly to the three-point correlation numbers, we encounter some problems with fulfilling the selection rules. Namely, the integral $\int dx Y_u^{k_1k_2k_4}Y_u^{k_3}$ is automatically zero when $k_1+k_2+k_3 > k_4 - 4$ because at this region the degree of the polynomial $Y_u^{k_1k_2k_4}$ is less than the degree of the polynomial $Y_u^{k_3}$. Thus it is not possible to satisfy (\ref{fusion 4-point 3s 2}) since in this case $k_1+k_2+k_3>p-2 \geq k_4 > k_4 -4$ and the integral automatically vanishes. 

However, as with the three-point correlation numbers, we can evaluate the four-point correlation numbers when the fusion rules are satisfied.
From the properties of the Jacobi polynomials we find that 

\be \label{Z^0}
 Z^{(0)}_{k_1k_2k_3k_4} = \begin{cases}
 -{1\over 4p^2}(2+k_1+k_2+k_3-k_4)(2p+3k_1+3k_2-3k_3-3k_4), \quad k_4 > k_1+k_2+k_3 \\
 -{1\over 2p^2}(1+k_1)(2p+3k_1+3k_2-3k_3-3k_4), \qquad\qquad\qquad k_1+k_4 \leq k_2+k_3
\end{cases}
\ee

With help of this and from the fusion rules (\ref{fusion 4-point 3s}) we find

\be \label{Qk1k2k3 2} \boxed{
Y_u^{k_1k_2k_4} = {1\over p^2} \sum_{k=0}^{k_4-k_1-k_2-s+p_0-6 \over 2} c_k x^{2(2-p_0)} P_k^{(0,{2\over 3}(3-2p_0))}(y) , \quad 
\begin{cases}
1\leq k_1,k_2 \leq s-1, \\ s \leq k_4 \leq p-2
\end{cases}
}
\ee 
where

\begin{align} 
&c_k = \nn \\
&={1\over 4}(6k+9-4p_0)(2+(k_1+k_2-k_4)+(2k+s-p_0+2))(2p+3(k_1+k_2-k_4)-3(2k+s-p_0+2))
\end{align}
Besides, from the expression (\ref{Z^0}) for $Z^{(0)}_{k_1k_2k_3k_4}$ when $k_1+k_4 \leq k_2+k_3$, we find that

\be \label{4-point normalized 2}
{ Z_{k_1k_2k_3k_4}Z_0 \over \left( \prod_{i=1}^4 Z_{k_ik_i} \right)^{1\over 2} } = {\prod_{i=1}^4 |p-3(k_i+1)|^{1\over 2} \over 2p(p+3)(p-3) } (1+k_1)(2p+3(k_1+k_2-k_3-k_4))
\ee
coincides with the result of direct calculation \cite{Belavin:2006ex} in Minimal Gravity (\ref{4-point MG}) where $m_i=1, n_i=k_i+1, q=3$.

$\bullet$ {$1\leq k_1\leq s-1, s\leq k_2,k_3,k_4 \leq p-2$; $p,k_i$ -- even}

The 4-point numbers are singular (\ref{dimension 4-point}). The expression for them reduces to 

\begin{align}
Z_{k_1k_2k_3k_4}^{(0)} =  & \left( - {(Y_u^0)'' \over ((Y_u^0)')^3}  + {\sum_{i=1}^4 (Y_u^{k_i})' \over ((Y_u^0)')^2} -{\sum_{i=2}^4 Y_u^{k_1k_i} \over (Y_u^0)'} \right)\Bigg|_{x=1}  \\
Z_{k_1k_2k_3k_4}^{(I)} = & \int_0^1 dx (Y^{k_2k_3k_4}_u Y^{k_1}_u + Y^{0}_u Y^{k_1k_2k_3k_4}_u )
\end{align}
We find that 

\be \label{4-point normalized 3}
{ Z^{(0)}_{k_1k_2k_3k_4}Z_0 \over \left( \prod_{i=1}^4 Z_{k_ik_i} \right)^{1\over 2} } = {\prod_{i=1}^4 |p-3(k_i+1)|^{1\over 2} \over 2p(p+3)(p-3) } (1+k_1)(-4p+3+3(-k_1+k_2+k_3+k_4))
\ee
This coincides with the direct calculation in Minimal Gravity (\ref{4-point MG}).

$\bullet$ {$s\leq k_1,k_2,k_3,k_4 \leq p-2$; $p,k_i$ -- even}

The 4-point numbers are non-singular if $\sum_i k_i \geq 2(p-1)$ (\ref{dimension 4-point}). The expression for them reduces to

\begin{align} 
Z_{k_1k_2k_3k_4}^{(0)} =  & \left( - {(Y_u^0)'' \over ((Y_u^0)')^3}  + {\sum_{i=1}^4 (Y_u^{k_i})' \over ((Y_u^0)')^2} \right)\Bigg|_{x=1}   \\
Z_{k_1k_2k_3k_4}^{(I)} = & \int_0^1 dx (Y^{k_1k_2k_3}_u Y^{k_4}_u + Y^{k_1k_2k_4}_u Y^{k_3}_u + Y^{k_1k_3k_4}_u Y^{k_2}_u + Y^{k_2k_3k_4}_u Y^{k_1}_u + Y^{0}_u Y^{k_1k_2k_3k_4}_u )
\end{align}   
We find that 
\be \label{4-point normalized 4}
{ Z^{(0)}_{k_1k_2k_3k_4}Z_0 \over \left( \prod_{i=1}^4 Z_{k_ik_i} \right)^{1\over 2} } = {\prod_{i=1}^4 |p-3(k_i+1)|^{1\over 2} \over 2p(p+3)(p-3) } ({3\over 2} \sum_{i=1}^4k_i^2  +(3-p)\sum_{i=1}^4k_i +{p^2\over 2} -4p+4 )
\ee
which again coincides with the direct calculation in Minimal Gravity under appropriate opening of the module in(\ref{4-point MG}).

In the last two cases we also did not manage to obtain a full agreement with the Liouville Minimal Gravity if the fusion rules are violated, however the value of the four-point correlation numbers coincide with \ref{4-point MG} when the fusion rules are satisfied.

Summarizing the results we see that: {\it The results (\ref{3-point MG}), (\ref{4-point MG}) of the direct calculations of the correlation numbers in Minimal Liouville 
Gravity are reproduced by the Douglas equation approach when the fusion rules are satisfied.  }

\section{Conclusion}

To sum up, we have considered the Douglas approach to Liouville Minimal Gravity. We suggested a convenient integral representation (\ref{free energy integral}) for the partition function of the Liouville Minimal Gravity perturbed by primary operators (\ref{observables}). To find the correlation numbers from this representation one needs to know two more things: an appropriate solution of the Douglas string equation and the resonance relations. In the case of $(3,3s+p_0)$ Liouville Minimal Gravity we proposed such solution of the Douglas string equation. Using this input we managed to satisfy all conformal and fusion selection rules in $(3,3s+p_0)$ Liouville Minimal Gravity up to two-point correlation numbers. 
In three-point correlation numbers we found that not all of the selection rules can be satisfied. The similar situation is found in the four-point correlation numbers. However, when the selection rules are satisfied, the three- and four-point correlation numbers (\ref{normalized three-point}), (\ref{4-point normalized 1}),  (\ref{4-point normalized 5}), (\ref{4-point normalized 2}), (\ref{4-point normalized 3}), (\ref{4-point normalized 4}), derived from the Douglas string equation approach for even $p$ and physical operators $O_{1,k}$ with odd $k$, coincide with the results of the direct calculations in $(q,p)=(3,3s+p_0)$ Minimal Liouville Gravity (\ref{3-point MG}), (\ref{4-point MG}).

The problem of satisfying the selection rules remains open. It seems to be similar to the paper \cite{Zamolodchikov:2005sj} where the correlation numbers of generalized Liouville Minimal Gravity were calculated and found to produce some finite numbers when the selection rules are violated. We leave this problem for the future.

\section*{Acknowledgements}

The authors thank V.Belavin, M. Bershtein, G. Delfino, M. Lashkevich, Y. Pugai and A.Zamolodchikov for useful discussions. The work of A.B. and B.M. was supported by RFBR grant no.13-01-90614 and by the Russian Ministry of Education and Science under the grants no.8528 and no.8410.

The work of B.D. was partially supported by the European Research Council Advanced Grant FroM-PDE, by the Russian Federation Government Grant No. 2010-220-01-077 and by PRIN 2010-11 Grant ``Geometric and analytic theory of Hamiltonian systems in finite and infinite dimensions'' of Italian Ministry of Universities and Researches.

\newpage

{\appendix

\section{Frobenius manifolds}

In this Appendix we will outline some basic ideas of the theory of Frobenius manifolds following \cite{Dubrovin:1992dz}, \cite{Dubrovin1996}.

\subsection{Basic definitions. Deformed flat coordinates on a Frobenius manifold}

An $n$-dimensional commutative associative algebra
$$
A={\rm span} (e_1, \dots, e_n)
$$
with a unit equipped with a nondegenerate symmetric \emph{invariant} bilinear form $(~,~)$ is called \emph{Frobenius algebra}. Invariance of the bilinear form means that the multiplication operators in the algebra are symmetric with respect to the bilinear form, i.e., the following identity
\begin{equation}\label{inva}
(a\cdot b, c)=(a, b\cdot c)
\end{equation}
holds true for any triple of vectors in $A$. It will be convenient to choose a basis $e_1$, \dots, $e_n$ in such a way that the vector $e_1$ coincides with the unit. Denote
\begin{equation}\label{eta}
\eta_{\alpha\beta}=(e_{\alpha}, e_{\beta})
\end{equation}
the Gram matrix of the bilinear form and $c_{\alpha\beta}^{\gamma}$ the structure constants of the algebra,
$$
e_{\alpha}\cdot e_{\beta} =c_{\alpha\beta}^{\gamma} e_{\gamma}.
$$
They satisfy
$$
c_{1\alpha}^{\beta} =\delta_{\alpha}^{\beta}.
$$
Moreover, lowering the upper index one obtains a totally symmetric 3-tensor
$$
c_{\alpha\beta\gamma}:= \eta_{\gamma\rho} c_{\alpha\beta}^{\rho}.
$$

Let us now consider an $n$-parameter family of $n$-dimensional Frobenius algebras $A_{v^1, \dots, v^n}$ of the above form satisfying the following properties.

1. The vector $e_1$ is the unit for all algebras $A_{\bf v}$ (here and below we will denote by ${\bf v}$ the entire vector of parameters ${\bf v}=(v^1, \dots, v^n)$.)

2. The Gram matrix of the invariant bilinear form on the Frobenius algebra $A_{\bf v}$ is ${\bf v}$-independent,
$$
(e_{\alpha}, e_{\beta})\equiv \eta_{\alpha\beta}={\rm const}.
$$

The dependence on ${\bf v}$ of the structure constants $c_{\alpha\beta}^{\gamma} ({\bf v})$ of the family of algebras is constrained by the following main condition.

3. There exists a function $F=F({\bf v})$ such that
\begin{equation}\label{triple}
c_{\alpha\beta\gamma}({\bf v}) =\frac{\partial^3 F({\bf v})}{\partial v^{\alpha} \partial v^{\beta} \partial v^{\gamma}}, \quad \alpha, \, \beta, \, \gamma=1, \dots, n.
\end{equation}
Associativity of the algebras implies the following system of nonlinear PDEs for the function $F$
\begin{equation}\label{wdvv}
\frac{\partial^3 F({\bf v})}{\partial v^{\alpha} \partial v^{\beta} \partial v^{\rho}} \eta^{\rho\lambda} \frac{\partial F({\bf v})}{\partial v^{\lambda} \partial v^{\mu} \partial v^{\nu}}=\frac{\partial^3 F({\bf v})}{\partial v^{\nu} \partial v^{\beta} \partial v^{\rho}} \eta^{\rho\lambda} \frac{\partial F({\bf v})}{\partial v^{\lambda} \partial v^{\mu} \partial v^{\alpha}}, \quad \alpha, \, \beta, \, \mu, \, \nu=1, \dots, n
\end{equation}
called \emph{WDVV associativity equations}. The system \eqref{wdvv} is overdetermined for $n\geq 4$. The function $F$ also satisfies
\beq\label{wdvv1}
\frac{\p^3 F}{\p v^1 \p v^{\alpha} \p v^{\beta}}=\eta_{\alpha\beta}
\eeq
due to the normalization $\p / \p v^1=$unit and constancy of the metric.

4. In all main examples  $F({\bf v})$ is a quasihomogeneous function  of the variables $v^{\alpha}$ or $\exp v^{\beta}$ of some degrees $(1-q_{\alpha})$ or $r_{\beta}$ respectively for some constants $q_1$, \dots, $q_n$, $r_1$, \dots, $r_n$ such that
\begin{equation}\label{quasi}
\sum_{\alpha=1}^n (1-q_{\alpha}) v^{\alpha}\frac{\partial F}{\partial v^{\alpha}} +\sum_{\beta=1}^n r_{\beta} \frac{\partial F}{\partial v^{\beta}}= (3-d)F +\mbox{terms at most quadratic in}~ v
\end{equation}
for some constant $d$.
Here the numbers $q_{\alpha}$ and $r_{\beta}$ satisfy
$$
q_1=0, \quad r_{\beta}\neq 0 \quad \mbox{only if}\quad q_{\alpha}=1.
$$
Validity of such a quasihomogeneity property will also be assumed. Under this assumption the WDVV associativity equations \eqref{wdvv} reduce to a system of ODEs.

The notion of \emph{Frobenius manifolds} is a reformulation of the above properties 1--4 in terms of a geometrical structure on the space of parameters $v^1, \dots, v^n$. Namely, we consider these parameters as coordinates on an $n$-dimensional manifold $M$. The algebra $A_{\bf v}$ is identified with the tangent space $T_{\bf v}M$ at the point ${\bf v}=\left( v^1, \dots, v^n\right)$ by identifying the bases
$$
e_{\alpha} \leftrightarrow \frac{\partial}{\partial v^{\alpha}}, \quad \alpha=1, \dots, n.
$$
In other words, on the Frobenius manifold $M$ we can multiply tangent vectors at any point ${\bf v}\in M$ and obtain another tangent vector at the same point,
$$
\frac{\partial}{\partial v^{\alpha}}\cdot \frac{\partial}{\partial v^{\beta}}=c_{\alpha\beta}^{\gamma}({\bf v}) \frac{\partial}{\partial v^{\gamma}}.
$$
The bilinear form $(~,~)$ is now defined on the tangent spaces to $M$, so it can be considered as a metric on $M$
\begin{equation}\label{metro}
ds^2 =\eta_{\alpha\beta} dv^{\alpha} dv^{\beta}
\end{equation}
(not necessarily positive definite). Constancy of the Gram matrix means that the metric is \emph{flat}, i.e., its curvature identically vanishes. The parameters $v^1$, \dots, $v^n$ of the family of Frobenius algebras are now identified with the flat coordinates of the metric. Recall that flat coordinates are determined uniquely up to an affine transformation
$$
v^{\alpha} \mapsto A^{\alpha}_{\beta} v^{\beta} + b^{\beta}.
$$
Other systems of curvilinear coordinates on $M$ sometimes prove to be useful. In these curvilinear coordinates the metric is not constant anymore. The structure constants of the Frobenius algebras in a system of curvilinear coordinates ${\bf u}=(u^1, \dots, u^n)$, $u^{\alpha}=u^{\alpha}({\bf v})$, become covariant derivatives of the potential $F$,
$$
c_{\alpha\beta\gamma}({\bf u}):=\left( \frac{\partial}{\partial u^{\alpha}} \cdot \frac{\partial}{\partial u^{\beta}}, \frac{\partial}{\partial u^{\gamma}}\right)=\nabla_{\alpha}\nabla_{\beta}\nabla_{\gamma} F.
$$
Here $\nabla$ is the Levi-Civita connection for the metric \eqref{metro}. The covariant derivatives commute due to flateness of the metric.
Observe that (local) existence of a potential $F$ in curvilinear coordinates  is ensured by symmetry of the 4-tensor
\begin{equation}\label{fro4}
\nabla_{\delta} c_{\alpha\beta\gamma}({\bf u})=\nabla_{\gamma} c_{\alpha\beta\delta}({\bf u}).
\end{equation}
 Such a symmetry can be used as one of the main axioms in the coordinate-free definition of Frobenius manifold.

Finally, two vector fields $e$ and $E$ are also defined on a Frobenius manifold. One of them is the unit of the algebra $A_{\bf v}=T_{\bf v}M$,
$$
e=e_1=\frac{\partial}{\partial v^1}.
$$ 
Another one is the generator of scaling transformations entered into the quasihomogeneity property
$$
E=\sum_{\alpha=1}^n (1-q_{\alpha}) v^{\alpha}\frac{\partial }{\partial v^{\alpha}} +\sum_{\beta=1}^n r_{\beta} \frac{\partial }{\partial v^{\beta}}.
$$

Associativity of the algebras along with the symmetry \eqref{fro4} can be easily derived from vanishing of the curvature
of the following deformation $\tilde \nabla=\tilde\nabla(z)$ of the Levi-Civita connection
\begin{equation}
\tilde \nabla_a b =\nabla_a b +z\, a\cdot b.
\end{equation}
Here $a$, $b$ are vector fields on $M$, $z$ is a parameter of the deformation. In the coordinates $v^{\alpha}$ the covariant derivative of a vector field $X^{\alpha}$ is defined by the formula
$$
\tilde \nabla_{\alpha} X^{\gamma} =\frac{\p X^{\gamma}}{\p v^{\alpha}} +z\, c_{\alpha\beta}^{\gamma}({\bf v}) X^{\beta}. 
$$
The curvature of the connection must vanish identically in $z$
\begin{equation}\label{kurva}
\left[\tilde\nabla_{\alpha} (z), \tilde\nabla_{\beta}(z)\right]=0.
\end{equation} 
Such a \emph{deformed flat connection} is one of the main playing characters\footnote{The quasihomogeneity axiom $$\partial_E F =(3-d) F+\mbox{quadratic terms}$$ can also be reformulated in a geometric way in terms of an extension of the deformed flat connection in the direction of the parameter $z$. Here we will not use such an extension, see \cite{Dubrovin1996} for details.} of the theory of Frobenius manifolds. One of immediate consequences of flateness of $\tilde\nabla$ is

{\bf Lemma A.1}. {\it There exist $n$ formal series 
\begin{equation}\label{theta}
\theta^{\alpha}({\bf v}, z)=\sum_{k=0}^\infty \theta^{\alpha}_{k}({\bf v})z^k , \quad \alpha=1, \dots, n
\end{equation}
satisfying
\begin{equation}\label{defo}
\tilde\nabla(z) d\theta^{\alpha}({\bf v},z)=0
\end{equation}
and normalized by conditions
\begin{equation}\label{norm1}
\theta^{\alpha}_{0}({\bf v})=v^{\alpha}
\end{equation}
and
\begin{equation}\label{norm2}
\left( \nabla \theta^{\alpha}({\bf v}, -z), \nabla\theta^{\beta}({\bf v}, z)\right)\equiv \eta^{\alpha\beta}.
\end{equation}
}

The functions $\theta^1({\bf v}, z)$, \dots, $\theta^n({\bf v}, z)$ can be considered as flat coordinates for the connection $\tilde \nabla(z)$. In these coordinates the covariant derivatives with respect to the connection coincide with partial derivatives. So we will call them \emph{deformed flat coordinates}. The equation \eqref{defo} can be spelled out as the following system of $z$-dependent differential equations
\begin{equation}\label{defo1}
\frac{\p^2 \theta^{\lambda}}{\p v^{\alpha} \p v^{\beta}} =z\, c_{\alpha\beta}^{\gamma}({\bf v}) \frac{\p \theta^{\lambda}}{\p v^{\gamma}}.
\end{equation}
A system of $n$ linearly independent solutions of the system \eqref{defo1} provides one with a system of deformed flat coordinates. Their gradients satisfy
$$
\nabla \left( \nabla \theta^{\alpha}({\bf v}, z_1), \nabla \theta^{\beta}({\bf v}, z_2)\right) =(z_1+z_2) \nabla \theta^{\alpha}({\bf v}, z_1) \cdot \nabla\theta^{\beta}({\bf v}, z_2).
$$
So the normalization \eqref{norm2} can be achieved by taking a suitable $z$-dependent linear combination.

From \eqref{defo1} one arrives at the following recursion relation 
\begin{equation}\label{defo2}
\frac{\p^2 \theta^{\lambda}_{k+1}}{\p v^{\alpha} \p v^{\beta}} = c_{\alpha\beta}^{\gamma}({\bf v}) \frac{\p \theta^{\lambda}_k}{\p v^{\gamma}}
\end{equation}
With the help of this recursion the coefficients of the $z$-expansion of deformed flat coordinates can be determined by quadratures.

In sequel we will often use quantities with lower indices
\beq\label{thetal}
\theta_\alpha({\bf v}, z)=\eta_{\alpha\beta} \theta^\beta({\bf v}, z).
\eeq
Their coefficients of expansion
\beq\label{thetal1}
\theta_\alpha({\bf v}, z) =\sum_{k=1}^\infty \theta_{\alpha,k}({\bf v}) z^k
\eeq
satisfy same recursion \eqref{defo2}.

\subsection{An example: 2D TFT of the $A_n$ type and Frobenius structure on the space of polynomials of degree $n+1$}\label{secC2}

First nontrivial (i.e., with non-constant $c_{\alpha\beta}^{\gamma}({\bf v})$) examples of Frobenius manifolds appeared in \cite{Dijkgraaf:1990dj} in the framework of two-dimensional topological field theory (2D TFT). In this case the Frobenius manifold structure is realized on the space of polynomials $Q(p)$ of the form
\beq\label{an}
M=\left\{ Q(p)=p^{n+1} + u_1 p^{n-1} +\dots + u_n \right\}.
\eeq
Here the coefficients are arbitrary numbers. They can be considered as coordinates on the manifold. The tangent space to $M$ at any point can be identified with the space of polynomials of degree less or equal than $(n-1)$. Namely, a tangent vector at the point $Q(p)\in M$  is identified with the derivative of the polynomial along this vector. So, for example,
$$
\frac{\p}{\p u_{\alpha} } \leftrightarrow p^{n-\alpha}.
$$
The algebra structure on the tangent space at the point $Q(p)$ is obtained by identifying this space with the quotient algebra
$$
T_{Q(p)}M = {\mathbb C}[p]/(Q'(p)).
$$
Here $Q'(p)=\frac{dQ(p)}{dp}$. The inner product of two tangent vectors $\p_1$, $\p_2\in T_{Q(p)}M$ is defined by the residue pairing
\begin{equation}\label{resp}
\left( \p_1, \p_2\right)_{Q(p)}=-(n+1)\ress_{p=\infty} \frac{\p_1 Q (p) \p_2 Q(p)}{Q'(p)} dp.
\end{equation}
The coefficients of the polynomial are not flat coordinates coordinates for this metric. The flat coordinates are given by the following procedure. Consider the inverse function $p=p(Q)$ to the polynomial $Q(p)$ written as a Laurent series in powers of $z=Q^{\frac1{n+1}}$
\beq\label{plam}
p=z-\frac{1}{n+1}\left( \frac{v_1}{z}+\frac{v_2}{z^2}+\dots +\frac{v_n}{z^n}\right) +{\mathcal O}\left( \frac1{z^{n+2}}\right).
\eeq
The first $n$ coefficients are flat coordinates with the Gram matrix
\beq\label{saito}
\eta^{\alpha\beta}=\left( \frac{\p}{\p v_{\alpha}}, \frac{\p}{\p v_{\beta}}\right) =  \delta_{\alpha+\beta, n+1}.
\eeq
So, raising and lowering indices follows the rule
$$
v^{\alpha}=v_{n-\alpha+1}.
$$
In these coordinates the structure constants of the Frobenius algebras are defined by the following residue
$$
c_{\alpha\beta\gamma}=\frac{\partial^3 F}{\partial v^{\alpha} \partial v^{\beta} \partial v^{\gamma}}=-(n+1)\ress_{p=\infty}\frac{\frac{\p Q(p)}{\p v^{\alpha}} \frac{\p Q(p)}{\p v^{\beta}} \frac{\p Q(p)}{\p v^{\gamma}}}{Q'(p)} dp.
$$
Finally we give a formula for the deformed flat coordinates (we will work with thetas with lower indices, see \eqref{thetal} above). Their coefficients $\theta_{\alpha,k}$ are given by the following residues
\beq\label{theta2}
\theta_{\alpha,k}= -c_{\alpha,k} \ress_{p=\infty} Q^{k+\frac{\alpha}{n+1}}(p) dp
\eeq
where
\beq\label{koef}
c_{\alpha,k} =\left[ \frac{\alpha}{n+1}\left( \frac{\alpha}{n+1}+1\right)\dots \left(\frac{\alpha}{n+1}+k\right)\right]^{-1}.
\eeq

In order to derive the above statements let us first prove the formula
\begin{equation}\label{phi}
\phi_{\alpha}(p):= \frac{\partial Q(p)}{\partial v^{\alpha}} =\frac{d}{dp} \frac1{\alpha}\, \left( Q^{\frac{\alpha}{n+1}}(p)\right)_+.
\end{equation}
Here the symbol $(~)_+$ is used for the polynomial part of the Laurent series. In order to prove the formula the following ``thermodynamic identity" will be useful
\begin{equation}\label{therm}
\frac{\partial}{\partial t} \left( Q(p) dp\right)_{p={\rm const}}=-\frac{\partial}{\partial t}\left( p(Q)dQ\right)_{Q={\rm const}}
\end{equation}
valid for any function $Q(p)$ depending on a parameter $t$.
Using this identity along with the definition \eqref{plam} we obtain
$$
\frac{\partial}{\partial v^{\alpha}} Q(p) dp = \left[z^{\alpha-1}+{\mathcal O}\left(z^{-2}\right)\right]\, dz=d\left[ \frac{z^{\alpha}}{\alpha} +{\mathcal O}\left( \frac1{z}\right)\right].
$$
This proves \eqref{phi}. 
In particular,
$$
\phi_1=1, \quad \phi_2=p.
$$

Now, the metric \eqref{saito} readily follows:
\begin{eqnarray}
&&
\left( \frac{\partial}{\partial v^{\alpha}}, \frac{\partial}{\partial v^{\beta}}\right) =-(n+1)\ress_{p=\infty} \frac{\frac{\partial Q(p)dp}{\partial v^{\alpha}} \frac{\partial Q(p)dp}{\partial v^{\beta}} }{dQ(p)}
\nonumber\\
&&
\nonumber\\
&&
=-\ress_{z=\infty} \frac{\left[ z^{\alpha-1}+{\mathcal O}\left(z^{-2}\right)\right] 
\, \left[ z^{\beta-1}+{\mathcal O}\left(z^{-2}\right)\right]}{z^n}\, dz=\delta_{\alpha+\beta, n+1}.
\nonumber
\end{eqnarray}
In a similar way one can derive an expression for the first derivatives of the potential of the structure constants, i.e., such function $F=F(v)$ that
$$
\frac{\partial^3 F}{\partial v^{\alpha} \partial v^{\beta} \partial v^{\gamma}} =c_{\alpha\beta\gamma}(v) =-(n+1)\ress_{p=\infty} \frac{\phi_{\alpha}(p) \phi_{\beta}(p) \phi_{\gamma}(p)}{Q'(p)}\, dp.
$$
To this end let us first prove that the functions $\theta_{\alpha,1}(v)$ (see  \eqref{theta2}) satisfy
\begin{equation} \label{h0}
\frac{\partial \theta_{\alpha,1}(v)}{\partial v^{\beta}}=\frac{\partial \theta_{\beta,1}(v)}{\partial v^{\alpha}}.
\end{equation}
Indeed, using \eqref{phi} one obtains
\begin{eqnarray}
&&
-\frac{\partial \theta_{\alpha,1}(v)}{\partial v^{\beta}}=\frac{n+1}{\alpha\, \beta} \ress Q^{\frac{\alpha}{n+1}}(p) \, d\left[ Q^{\frac{\beta}{n+1}}(p)\right]_+=\frac{n+1}{\alpha\, \beta} \ress \left[Q^{\frac{\alpha}{n+1}}(p)\right]_- \, d\left[ Q^{\frac{\beta}{n+1}}(p)\right]_+
\nonumber\\
&&
=-\frac{n+1}{\alpha\, \beta} \ress d\left[Q^{\frac{\alpha}{n+1}}(p)\right]_- \, \left[ Q^{\frac{\beta}{n+1}}(p)\right]_+=-\frac{n+1}{\alpha\, \beta} \ress d\left[Q^{\frac{\alpha}{n+1}}(p)\right]_- \, \left[ Q^{\frac{\beta}{n+1}}(p)\right]
\nonumber\\
&&
=\frac{n+1}{\, \alpha\beta} \ress d\left[Q^{\frac{\alpha}{n+1}}(p)\right]_+ \, \left[ Q^{\frac{\beta}{n+1}}(p)\right]=-\frac{\partial \theta_{\beta,1}(v)}{\partial v^{\alpha}}.
\nonumber
\end{eqnarray}
Therefore there exists a function $F(v)$ (a polynomial) such that
\begin{equation}\label{df}
\frac{\partial F}{\partial v^{\alpha}} =\theta_{\alpha,1}(v)= -\frac1{\frac{\alpha}{n+1}\left( \frac{\alpha}{n+1} +1\right)} \ress_{p=\infty} Q^{1+\frac{\alpha}{n+1}}(p)\, dp.
\end{equation}
(cf. formula (4.47) in \cite{Dijkgraaf:1990dj}). 

We postpone till a subsection \ref{subC4} the proof of the recursion relations \eqref{defo2} for the functions $\theta_{\alpha,p}$.

For the case $n=2$ the coefficients of the polynomial $Q(p)=p^3+u_1 p+u_2$ turn out to be flat coordinates,
$$
u_1=v^2, \quad u_2=v^1.
$$
Like above redenote
$$
u_1\mapsto u, \quad u_2\mapsto v.
$$
The multiplication table in the Frobenius algebra reduces to only one non obvious formula
$$
e_2\cdot e_2= -\frac13 u \,e_1.
$$
So the potential of the Frobenius manifold reads
$$
F=\frac12 v^2 u -\frac{u^4}{72}.
$$

The first few coefficients of expansion of deformed flat coordinates read
\begin{eqnarray}
&&
\theta_{1,0}=u, \quad \theta_{2,0}=v
\nonumber\\
&&
\theta_{1,1}=u\, v, \quad \theta_{2,1}=\frac12 v^2 -\frac1{18} u^3
\nonumber\\
&&
\theta_{1,2}=\frac12 u\, v^2 -\frac1{36} u^4, \quad \theta_{2,2}=\frac16 v^3 -\frac1{18} u^3 v.
\nonumber
\end{eqnarray}

{\bf Remark A.2}. \label{cotan} Due to nodegeneracy of the metric one can define a structure of Frobenius algebra also on the cotangent spaces $T_{{\bf v}}^*M$,
$$
(dv^{\alpha}, dv^{\beta})=\eta^{\alpha\beta}
$$
(the inverse matrix to $\eta_{\alpha\beta}$) and
\begin{equation}\label{dvdv}
dv^{\alpha}\cdot dv^{\beta} = c^{\alpha\beta}_{\gamma}({\bf v}) dv^{\gamma}\quad \mbox{where} \quad c^{\alpha\beta}_{\gamma} =\eta^{\alpha\delta} c_{\delta\gamma}^{\beta}.
\end{equation}
In the case of the space of polynomials this structure can be described by the following representations
\begin{eqnarray}
&&
(dQ(p), dQ(q))\equiv \sum\limits_{\alpha,\beta=1}^n (du_{\alpha} ,du_{\beta} ) p^{n-\alpha} q^{n-\beta}= -\frac1{n+1}\frac{Q'(q)-Q'(p)}{p-q}
\nonumber\\
&&
dQ(p)\cdot dQ(q) \equiv \sum\limits_{\alpha,\beta=1}^n du_\alpha \cdot du_\beta \,p^{n-\alpha} q^{n-\beta}= \frac{Q'(q) dQ(p) -Q'(p) dQ(q)}{p-q}.
\nonumber
\end{eqnarray}
In the example $n=2$ the multiplication of 1-forms is defined by the following table
$$
du\cdot du=du, \quad du\cdot dv=dv, \quad dv\cdot dv=-\frac{u}3 du.
$$

\medskip

\subsection{From Frobenius manifolds to integrable hierarchies}

With any Frobenius manifold $M$ we will now associate an infinite dimensional space ${\cal M}$ equipped with a Poisson bracket and with an infinite family of pairwise commuting Hamiltonians. The space ${\cal M}$ is defined as the space of smooth functions of one variable $x$ with values in $M$. For two functionals $I$ and $J$ on this space define their Poisson bracket by the formula
$$
\{ I, J\} =\int \frac{\delta I}{\delta v^{\alpha}(x)} \eta^{\alpha\beta} \frac{d}{dx} \frac{\delta J}{\delta v^{\beta}(x)} \, dx.
$$
Here the integral is defined like in the formal variational calculus, i.e., considering the class of equivalence of the integrand modulo total derivatives. 

In order to construct a family of commuting Hamiltonians we will use the coefficients of expansion of deformed flat coordinates (see above Lemma C.1)

{\bf Lemma A.3}. {\it The functionals
$$
H_{\alpha,k} =\int \theta_{\alpha,k+1}({\bf v})\, dx, \quad \alpha=1, \dots, n, \quad k\geq 0
$$
commute pairwise,
$$
\{ H_{\alpha,k}, H_{\beta,l}\}=0.
$$
}

Recall that functions $\theta_{\alpha,k}({\bf v})$ are coefficients of expansion of deformed flat coordinates with lowered indices, see \eqref{thetal}, \eqref{thetal1}.

{\it Proof}. It suffices to prove commutativity
$$
\{ H_{\alpha}(z_1), H_{\beta}(z_2)\}=0
$$ 
of the generating series functionals
$$
H_{\alpha}(z_1) =\int \theta_{\alpha}({\bf v}, z_1)\, dx, \quad H_{\beta}(z_2) =\int \theta_{\beta}({\bf v}, z_2)\, dx.
$$
We have
\begin{align*}
&
\{ H_{\alpha}(z_1), H_{\beta}(z_2)\} =\int \frac{\p \theta_{\alpha}({\bf v}, z_1)}{\p v^{\gamma}} \eta^{\gamma\tau}  \frac{\p^2 \theta_{\beta}({\bf v}, z_2)}{\p v^{\lambda}\p v^{\tau}} v^{\lambda}_x \, dx=z_2\int \frac{\p \theta_{\alpha}({\bf v}, z_1)}{\p v^{\gamma}} c^{\gamma\tau}_{\lambda} \frac{\p \theta_{\beta}({\bf v}, z_2)}{\p v^{\tau}} v^{\lambda}_x\, dx\\
&
=z_2\int \left[ d\theta_{\alpha}({\bf v}, z_1) \cdot d\theta_{\beta}({\bf v}, z_2)\right]_{\lambda} v^{\lambda}_x \, dx
\end{align*}
where the product of one-forms is defined by the Frobenius algebra structure on the cotangent space $T_{\bf v}^*M$ (see Remark C.2 above). To complete the proof of commutativity it remains to prove that the one-form $d\theta_{\alpha}({\bf v}, z_1) \cdot d\theta_{\beta}({\bf v}, z_2)$ is closed. To this end consider the function
\begin{equation}\label{omega}
\Omega_{\alpha\beta}({\bf v}, z_1, z_2) =\frac{(\nabla \theta_{\alpha}({\bf v}, z_1), \nabla \theta_{\beta}({\bf v}, z_2))-\eta_{\alpha\beta}}{z_1+z_2}.
\end{equation}
Due to normalization \eqref{norm1} the numerator vanishes at $z_2=-z_1$. So the right hand side of \eqref{omega} is a power series in $z_1$ and $z_2$. With the help of the defining equation \eqref{defo1} one can easily check that
\begin{equation}\label{omega1}
d\Omega_{\alpha\beta}({\bf v}, z_1, z_2) =  d\theta_\alpha({\bf v}, z_1) \cdot d\theta_\beta({\bf v}, z_2) \equiv c^{\tau\lambda}_{\rho}({\bf v}) \frac{\p \theta_{\alpha}({\bf v}, z_1)}{\p v^{\tau}}\frac{\p \theta_{\beta}({\bf v}, z_2)}{\p v^{\lambda}}dv^{\rho}
\end{equation}
(cf. eq. \eqref{dvdv}). This completes the proof of the Lemma.

\medskip

Due to commutativity of Hamiltonians we arrive at a family of pairwise commuting Hamiltonian flows
\begin{equation}\label{ham}
\frac{\p {\bf v}}{\p t^\alpha_k} = \{ {\bf v}(x), H_{\alpha,k}\} =\p_x \nabla \theta_{\alpha,k+1}({\bf v})=\nabla \theta_{\alpha,k}({\bf v})\cdot {\bf v}_x,
\end{equation}
$$
\frac{\p}{\p t^\beta_l} \frac{\p {\bf v}}{\p t^\alpha_k} = \frac{\p}{\p t^\alpha_k} \frac{\p {\bf v}}{\p t^\beta_l}.
$$
In particular
$$
\frac{\p {\bf v}}{\p t^1_0}={\bf v}_x
$$
(recall: the vector $\p / \p v^1$ coincides with the unit of the Frobenius algebra), so the time variable $t^1_0$ can be identified with $x$.

We will now derive an analogue of the Douglas string equation for the general solution to the family of commuting flows. 
Define a function $S_{\bf t} ({\bf v})$ on the Frobenius manifold depending on the infinite number of time variables ${\bf t}=( t^\alpha_k)_{\alpha=1, \dots, n, ~ k\geq 0}$ by
$$
S_{\bf t} ({\bf v})=\sum_{\alpha=1}^n \sum_{k\geq 0} t^\alpha_k \theta_{\alpha,k}({\bf v}).
$$
The ``string equation" will depend on an infinite number of arbitrary constants $c^i_p$ (an analogue of dilaton shift). Introduce shifted time variables $\tilde t^1_0 =t^1_0+x-c^1_0$ and
$$
\tilde t^\alpha_k =t^\alpha_k-c^{\alpha}_k
$$
for all other $1\leq \alpha\leq n$ and $k\geq 0$.
For a given choice of the shifts $c^{\alpha}_k$ consider a critical point ${\bf v}({\bf t})$ of  $S_{\tilde {\bf t}} ({\bf v})$ as function of the time variables assuming validity of the implicit function theorem for the resulting system of $n$ equations
\begin{equation}\label{krit}
\nabla S_{\tilde {\bf t}} ({\bf v})=0.
\end{equation}

{\bf Lemma A.4}. {\it For any choice of the shifts $c^{\alpha}_k$ the solution ${\bf v}({\bf t})$ satisfies the system of PDEs \eqref{ham}.}

{\it Proof}. Applying the implicit function theorem procedure we can compute the $x$- and $t$-derivatives of the function ${\bf v}({\bf t})$. To this end let us first differentiate the $\beta$-th equation of the system \eqref{krit}
$$
\sum \tilde t^\alpha_k \frac{\p \theta_{\alpha,k}}{\p v^{\beta}}=0
$$
over $x=t^1_0$. Obtain
$$
0=\frac{\p \theta_{1,0}}{\p v^{\gamma}}+\sum \tilde t^\alpha_k \frac{\p^2 \theta_{\alpha,k}}{\p v^{\beta} \p v^{\gamma}} v^{\beta}_x=\eta_{1\gamma} +\sum \tilde t^\alpha_k c_{\beta\gamma}^{\lambda} \frac{\p \theta_{\alpha,k-1}}{\p v^{\lambda}} v^{\beta}_x
$$
or, after raising the index $\gamma$
$$
\delta_1^{\gamma} +\sum \tilde t^\alpha_k c_{\beta}^{\gamma\lambda} \frac{\p \theta_{\alpha,k-1}}{\p v^{\lambda}} v^{\beta}_x=0.
$$
In this calculation we have used the normalization $\theta^{1}_{0}=v^1$ and the recursion \eqref{defo2}. The resulting system of equations can be rewritten in the vector form
\beq\label{doug1}
e+ \sum \tilde t^\alpha_k \nabla\theta_{\alpha,k-1} \cdot {\bf v}_x=0.
\eeq
So, the conditions of the implicit function theorem are fulfilled if the tangent vector $\sum \tilde t^\alpha_k \nabla\theta_{\alpha,k-1}$ is an invertible element of the Frobenius algebra. Under this assumption we obtain
$$
{\bf v}_x =- \left[ \sum \tilde t^\alpha_k \nabla\theta_{\alpha,k-1}\right]^{-1}
$$
(inversion in the Frobenius algebra).
A similar computation gives
$$
\frac{\p {\bf v}}{\p t^\beta_l} =- \nabla\theta_{\beta,l} \cdot  \left[ \sum \tilde t^\alpha_k \nabla\theta_{\alpha,k-1}\right]^{-1}.
$$
Hence
$$
\frac{\p {\bf v}}{\p t^\beta_l} =\nabla\theta_{\beta,l} \cdot {\bf v}_x.
$$
This proves that the solution to the least action principle equations \eqref{krit} satisfies the PDEs \eqref{ham}. The Lemma is proved. 

\medskip

We will now define a tau-function of the solution ${\bf v}({\bf t})$ constructed by the above procedure. Define the function ${\cal F}({\bf t})$ by the following formula
\begin{equation}\label{free}
{\cal F}({\bf t}) =\frac12 \left(\int dS_{\tilde{\bf t}} ({\bf v})\cdot dS_{\tilde{\bf t}} ({\bf v})\right)_{{\bf v}={\bf v}({\bf t})}= \frac12 \sum \tilde t^\mu_k \,\tilde t^\nu_l \left(\int c^{\alpha\beta}_\gamma({\bf v}) \frac{\p \theta_{\mu,k}({\bf v})}{\p v^\alpha}  \frac{\p \theta_{\nu,l}({\bf v})}{\p v^\beta} dv^\gamma\right)_{{\bf v}={\bf v}({\bf t})}.
\end{equation}
Here we use that the products of one forms $d\theta_{\mu,k}({\bf v}) \cdot d\theta_{\nu,l}({\bf v})=c^{\alpha\beta}_\gamma({\bf v}) \frac{\p \theta_{\mu,k}({\bf v})}{\p v^\alpha}  \frac{\p \theta_{\nu,l}({\bf v})}{\p v^\beta} dv^\gamma$ are closed one-forms (see above the proof of Lemma C.3).

{\bf Lemma A.5}. {\it The function \eqref{free} satisfies
$$
\frac{\p^2 {\cal F}({\bf t})}{\p t^\alpha_k \p t^1_0} =\theta_{\alpha,k}({\bf v}({\bf t})).
$$
In particular the components of the vector-valued function  ${ \bf v}({\bf t})$ are given by
$$
v^{\alpha}({\bf t}) =\eta^{\alpha\beta} \frac{\p^2 {\cal F}({\bf t})}{\p t^\beta_0 \p t^1_0}, \quad \alpha=1, \dots, n.
$$
}

{\it Proof}. We have
$$
\frac{\p {\cal F}}{\p t^\alpha_k} = \sum \tilde t^\beta_l d^{-1} \left( d\theta_{\beta,l}\cdot d\theta_{\alpha,k}\right)_{{\bf v}={\bf v}({\bf t})}+\frac12 \sum\tilde t^\beta_l  \tilde t^\gamma_r (d\theta_{\beta,l}({\bf v}) \cdot d\theta_{\gamma,r}({\bf v}) )_{\lambda} \frac{\p v^{\lambda}}{\p t^\alpha_k}\left|_{{\bf v}={\bf v}({\bf t})}.\right.
$$
The second term vanishes. Indeed, it is equal to 
$$
\frac12 \left( dS_{\tilde {\bf t}} \cdot dS_{\tilde {\bf t}} , \frac{\p {\bf v}}{\p t^\alpha_k}\right)_{{\bf v}={\bf v}({\bf t})}=0
$$
since $dS_{\tilde {\bf t}} ({\bf v})_{{\bf v}={\bf v}({\bf t})}=0$. Differentiating again and using similar arguments obtain
$$
\frac{\p^2 {\cal F}({\bf t})}{\p t^\alpha_k \p t^1_0} =d^{-1} \left( d\theta_{1,0} \cdot d\theta_{\alpha,k}\right)=\theta_{\alpha,k}
$$
since $d\theta_{1,0}=dv_1$ is the unit of the Frobenius algebra structure on the cotangent space $T_{\bf v}^*M$. The Lemma is proved.

\medskip

{\bf Definition A.6}. The function
\begin{equation}\label{tau}
\tau({\bf t})= e^{{\cal F}({\bf t})}
\end{equation}
is called the \emph{tau-function} of the solution \eqref{krit}.

\medskip

Notice that the primitives of the closed one-forms $d\theta_{\alpha,k}({\bf v}) \cdot d\theta_{\beta,l}({\bf v})$ can be defined by the generating function \eqref{omega},
\begin{equation}\label{omega11}
\sum_{k,l\geq 0} d^{-1}\left( d\theta_{\alpha,k}({\bf v}) \cdot d\theta_{\beta,l}({\bf v})\right) z_1^k z_2^l= \Omega_{\alpha\beta}({\bf v}, z_1, z_2).
\end{equation}
This fixes uniquely the integration constants in the definition of tau-function. After such a choice it is not difficult to prove validity of the \emph{string equation} for the logarithm of the tau-function.

\medskip

{\bf Lemma A.7}. {\it The function \eqref{free} normalised as in \eqref{omega11} satisfies the string equation
\begin{equation}\label{struna}
\frac12 \eta_{\alpha\beta} \tilde t^\alpha_0\, \tilde t^\beta_0 +\sum \tilde t^\alpha_{k+1} \frac{\p {\cal F}}{\p t^\alpha_k} =0.
\end{equation}
}

The proof is straightforward.

\medskip

{\bf Example A.8}. Consider a particular solution ${\bf v}({\bf t})$ specifying the shifts as follows
$$
c^1_1=1, \quad \mbox{all other}\quad c^{\alpha}_k=0.
$$
For this solution the least action eqs. become
$$
{\bf v}=\nabla S_{\bf t}({\bf v}).
$$
Restricting onto the \emph{small phase space}
$$
t^\alpha_k=0 \quad \mbox{for}\quad k>0
$$
one obtains
$$
v^{\alpha}=t^\alpha_0, \quad \alpha=1, \dots, n.
$$
For this solution the restriction of \eqref{free} onto the small phase space coincides with the potential of the Frobenius manifold
$$
{\cal F}({\bf t})|_{t^\alpha_0=v^{\alpha}, ~ t^\alpha_k=0~{\rm for}~ k>0}=F({\bf v}).
$$
For physically interesting examples of Frobenius manifolds the logarithm of the tau-function of this particular solution coincides with the tree-level free energy of the 2D TFT coupled with topological gravity.
}

\subsection{Frobenius manifold of $A_n$ type and dispersionless limit of the Gelfand--Dickey hierarchy}\label{subC4}

The Gelfand--Dickey (GD) hierarchy (it also coincides with the Drinfeld--Sokolov hierarchy of $A_n$ type) is an infinite sequence of pairwise commuting PDEs for $n$ functions $u_1(x,t)$, \dots, $u_n(x,t)$. Here $x$ is the spatial variable and $t$ is one of the time variables. The equations of the hierarchy are conveniently represented in the Lax form
\beq\label{gd1}
\frac{\p \hat{Q}}{\p t^{\alpha}_k}=\left[ \hat{A}_{\alpha,k}, \hat{Q}\right], \quad \alpha=1, \dots, n, \quad k\geq 0.
\eeq
Here the ordinary linear differential operators $\hat{Q}$ and $\hat{A}_{\alpha,k}$ are defined by
\beq\label{lax1}
\hat{Q}=\frac{d^{n+1}}{dx^{n+1}} + u_1(x) \frac{d^{n-1}}{dx^{n-1}}+\dots + u_n(x)
\eeq
\beq\label{lax2}
\hat{A}_{\alpha,k}=\frac1{n+1}c_{\alpha,k}\left( \hat{Q}^{k+\frac{\alpha}{n+1}}\right)_+ = c_{\alpha,k} \frac{d^k}{dx^k}+\dots.
\eeq
Here $\left( \hat{Q}^{k+\frac{\alpha}{n+1}}\right)_+$ denotes the differential part of the pseudodifferential operator $ \hat{Q}^{k+\frac{\alpha}{n+1}}$; the normalizing coefficient $c_{\alpha,k}$ is chosen as in \eqref{koef}. Equations of the hierarchy commute pairwise
$$
\frac{\p}{\p t^\beta_l}\frac{\p \hat{Q}}{\p t^\alpha_k}=\frac{\p }{\p t^\alpha_k}\frac{\p \hat{Q}}{\p t^\beta_l}.
$$
They can also be represented in the Hamiltonian form 
$$
\frac{\p u_{\beta}}{\p t^\alpha_k} ={\cal A}_{\beta\gamma} \frac{\delta H_{\alpha,k}}{\delta u_{\gamma}(x)}
$$
with a Hamiltonian operator ${\cal A}_{\beta\gamma}$
(see details in \cite{GD76}) with the Hamiltonians
$$
H_{\alpha,k}=c_{\alpha,k+1}\int dx \ress \hat{Q}^{k+1+\frac{\alpha}{n+1}}.
$$
Here $\ress$ is the M.Adler trace \cite{adler} defined as the coefficient of $\left(\frac{d}{dx}\right)^{-1}$ of the pseudodifferential operator $\hat{Q}^{k+1+\frac{\alpha}{n+1}}$.

We will now apply to the equations of GD hierarchy the procedure of dispersionless limit. In general the dispersionless limit of a PDE
$$
v_t = F(v, v_x, v_{xx}, \dots)
$$
satisfying $F(v, 0, 0, \dots)\equiv 0$
is obtained by applying a rescaling
$$
x\mapsto \epsilon x,\quad t\mapsto \epsilon t
$$
and then taking the limit of the resulting equation at $\epsilon\to 0$. In the case of the GD equations one can apply the following recipe. First, the differential operators have to be replaced by their symbols substituting an independent variable $p$ instead of $\frac{d}{dx}$. For example, the symbol of $\hat{Q}=\hat{Q}\left(\frac{d}{dx}\right)$ is our polynomial
$$
Q(p)=p^{n+1}+u_1 p^{n-1}+\dots +u_n.
$$
Notice that the symbol $A_{\alpha,k}$ of the operator $\hat{A}_{\alpha,k}$ is given by the polynomial part of the Laurent series
\beq\label{alphaiq}
A_{\alpha,k}(p)=\frac1{n+1}c_{\alpha,k}\left(Q^{k+\frac{\alpha}{n+1}}(p)\right)_+.
\eeq
Next, the commutators of the operators have to be replaced by the Poisson bracket of the symbols. So, the dispersionless limit of eq. \eqref{gd1} reads
\beq\label{dgd1}
\frac{\p Q(p)}{\p t^\alpha_k}=\left\{ A_{\alpha,k}, Q\right\}\equiv \frac{\p A_{\alpha,k}}{\p p}\frac{\p Q}{\p x}-\frac{\p Q}{\p p}\frac{\p A_{\alpha,k}}{\p x}.
\eeq
In the left hand side of this equation one has to differentiate in time the coefficients of the polynomial $Q(p)$. The independent variable $p$ is kept constant.
An alternative representation of equations of the hierarchy can be obtained by replacing the polynomial $Q(p)$ by the inverse function $p(Q)$. Now we want to differentiate in time the series representation \eqref{plam} of $p(Q)$ keeping $z=Q^{\frac1{n+1}}={\rm const}$. The resulting equation reads\beq\label{dgd2}
\frac{\p p(Q)}{\p t^{\alpha}_k} = \frac{\p }{\p x} (A_{\alpha,k})_{Q={\rm const}}.
\eeq
It is understood that, in the right hand side of this formula one has to differentiate in $x$ the coefficients of the Laurent expansion in $z=Q^{\frac1{n+1}}$ of 
$
A_{\alpha,k}\left(p(Q)\right).
$

Our goal is to prove that the dispersionless GD equation \eqref{dgd2} is equivalent to the Hamiltonian equation
\beq\label{ura1}
\frac{\p {\bf v}}{\p t^\alpha_k} =\p_x \nabla \theta_{\alpha,k+1}({\bf v})
\eeq
constructed in the previous section for an arbitrary Frobenius manifold. Here the functions $\theta_{\alpha,k}({\bf v})$ for the $A_n$ Frobenius manifold are defined by the formula \eqref{theta2}. Indeed, multiplying \eqref{dgd2} by $Q^{-\frac{\beta}{n+1}}\, dQ =(n+1) z^{n-\beta}dz$ and taking the residue at infinity we obtain
$$
\frac{\partial v^{\beta}}{\partial t^\alpha_k} =\frac{\partial}{\partial x} \ress Q^{-\frac{\beta}{n+1}} A_{\alpha,k}\, dQ.
$$
Let us compute the residue in the rhs. From the formula \eqref{phi} it immediately follows that the derivatives $\phi_{\alpha}(p)$ of the polynomial $Q(p)$ with respect to the flat coordinates coincide with the $p$-derivatives of $A_{\alpha,0}(p)$,
$$
\phi_{\alpha}(p)=\frac{d}{dp} A_{\alpha,0}(p).
$$
So, we have
\begin{eqnarray}
&&
(n+1)\ress Q^{-\frac{\beta}{n+1}} A_{\alpha,k}\, dQ=c_{\alpha,k}\ress Q^{-\frac{\beta}{n+1}}(p)\left( Q^{k+\frac{\alpha}{n+1}}(p)\right)_+ Q'(p)\, dp
\nonumber\\
&& 
=-c_{\alpha,k}\ress Q^{-\frac{\beta}{n+1}}(p)\left( Q^{k+\frac{\alpha}{n+1}}(p)\right)_- Q'(p)\, dp=-c_{\alpha,k}\ress \left(Q^{-\frac{\beta}{n+1}}(p)Q'(p)\right)_+\left( Q^{k+\frac{\alpha}{n+1}}(p)\right)_- \, dp
\nonumber\\
&&
=-c_{\alpha,k}\ress \left(Q^{-\frac{\beta}{n+1}}(p)Q'(p)\right)_+ Q^{k+\frac{\alpha}{n+1}}(p) \, dp=-c_{\alpha,k}\frac{n+1}{n-\beta+1} \ress \frac{d}{dp} \left(Q^{\frac{n-\beta+1}{n+1}}(p)\right)_+ Q^{k+\frac{\alpha}{n+1}}(p) \, dp
\nonumber\\
&&
=-(n+1) c_{\alpha,k} \ress \phi_{n-\beta+1} (p) Q^{k+\frac{\alpha}{n+1}}(p) \, dp=(n+1) \frac{\partial \theta_{\alpha,k+1}(v)}{\partial v^{n-\beta+1}}.
\nonumber
\end{eqnarray}
The formula \eqref{ura1} is proved. In particular, since
$$
\theta_{\alpha,1}(v)=\frac{\partial F}{\partial v^{\alpha}}
$$
(see the formula \eqref{df}), we have the following explicit form of the equations of the hierarchy \eqref{ura1} of the lowest level $k=0$
\begin{equation}\label{df1}
\frac{\partial v^{\gamma}}{\partial t^\alpha_0}=c_{\alpha\beta}^{\gamma}({\bf v})\frac{\partial v^{\beta}}{\partial x}.
\end{equation}

Using commutativity of the flows \eqref{ura1} with \eqref{df1} we derive
$$
c_{\alpha\beta}^{\gamma} \frac{\partial^2 \theta_{\alpha,k+1}}{\partial v^{\gamma} \partial v^{\lambda}} = c_{\alpha\lambda}^{\gamma}\frac{\partial^2 \theta_{\alpha,k+1}}{\partial v^{\beta} \partial v^{\gamma}}.
$$
Specializing at $\lambda=1$, since $c_{\alpha 1}^{\beta}=\delta_{\alpha}^{\beta}$, we arrive at
$$
\frac{\partial^2 \theta_{\alpha, k+1}}{\partial v^{\beta} \partial v^{\gamma}} = c_{\beta\gamma}^{\lambda} \frac{\partial^2 \theta_{\alpha,k+1}}{\partial v^{\lambda} \partial v^1}.
$$
With the help of an obvious identity
$$
\frac{\partial \theta_{\alpha,k+1}}{\partial v^1} =\theta_{\alpha,k}
$$
one arrives at the needed recursion relation
\begin{equation}\label{rec}
\frac{\partial^2 \theta_{\alpha,k+1}}{\partial v^{\beta} \partial v^{\gamma}} = c_{\beta\gamma}^{\lambda} \frac{\partial \theta_{\alpha,k}}{\partial v^{\lambda} }.
\end{equation}
Recall (see the previous section) that the equation \eqref{dgd2} can be recast into the form
\beq\label{dgd3}
\frac{\p {\bf v}}{\p t^\alpha_k}=\nabla\theta_{\alpha,k}({\bf v})\cdot {\bf v}_x
\eeq
(the product of tangent vectors on the Frobenius manifold).

\medskip

In the example $n=2$ the Lax operator reads
$$
\hat{Q}=\p_x^3 + u(x) \p_x +v(x)
$$
(we use the same notations $u_1\mapsto u$, $u_2\mapsto v$ as above). The first few flows of the GD hierarchy read
\begin{align*}
&
\frac{\p u}{\p t^1_0}=u_x\\
&
\frac{\p v}{\p t^1_0}=v_x
\end{align*}
\begin{align*}
&
\frac{\p u}{\p t^2_0} =\p_x\left( v-\frac12 u_x\right)\\
&
\frac{\p v}{\p t^2_0}=\p_x\left(-\frac16 u^2 + \frac12 v_x -\frac13 u_{xx}\right)
\end{align*}
Eliminating $v$ from this system one arrives at the Boussinesq equation
$$
u_{tt}+\frac16 \p_x^2 \left[ u^2 +\frac12 u_{xx}\right]=0, \quad t=t^2_0.
$$
Next,
\begin{align*}
&
\frac{\p u}{\p t^1_1}=\p_x \left( u\, v -\frac12 u\, u_x +\frac12 v_{xx} -\frac14 u_{xxx}\right)
\\
&
\frac{\p v}{\p t^1_1} =\p_x\left( -\frac19 u^3 +\frac12 v^2 -\frac14 u_x^2 +\frac12 u\, v_x -\frac12 u\, u_{xx} +\frac14 v_{xxx}-\frac16 u_{xxxx}\right)
\end{align*}
\begin{align*}
&
\frac{\p u}{\p t^2_1}=\p_x\left( -\frac1{18} u^3 +\frac12 v^2 -\frac12 v\, u_x -\frac16 u\, u_{xx} -\frac1{30} u_{xxxx}\right)
\\
&
\frac{\p v}{\p t^2_1}=\p_x\left( -\frac16 u^2 v +\frac12 v\, v_x -\frac13 u_x v_x -\frac13 v\, u_{xx} -\frac16 u\, v_{xx} -\frac1{30} v_{xxxx}\right)
\end{align*}
etc. One can easily check that the dispersionless limit of these equations has the needed form
\begin{align*}
&
\frac{\p u}{\p t^\alpha_k}=\p_x \frac{\p \theta_{\alpha,k+1}}{\p v}
\\
&
\frac{\p v}{\p t^\alpha_k}=\p_x \frac{\p \theta_{\alpha,k+1}}{\p u}
\end{align*}
where the functions $\theta_{\alpha,k}$ were constructed in Section \ref{secC2} above.

We will now show that, for the $A_n$ Frobenius manifold \eqref{an} the variational procedure \eqref{krit} for constructing solutions to the dispersionless GD hierarchy is equivalent to the (integrated in $x$) Douglas equation
\beq\label{doug0}
\{ Q, P\} =1
\eeq
with
$$
P=\sum\tilde t^\alpha_k A_{\alpha, k-1}(p).
$$
Indeed, according to \eqref{dgd3} the equation \eqref{doug1} obtained by differentiating eq. \eqref{krit} in $x$ implies that
$$
e+\sum\tilde t^\alpha_k \frac{\p {\bf v}}{\p t^\alpha_k}=0.
$$
Hence, for this solution we have
$$
\sum\tilde t^\alpha_k \frac{\p Q(p)}{\p t^\alpha_k}=-1.
$$
Taking into account the dispersionless Lax representation \eqref{dgd1} we arrive at eq. \eqref{doug0}.

\section{Jacobi polynomials}

Here we collect a few facts about Jacobi polynomials which we used throughout the text. The Jacobi polynomials $P_n^{(a,b)}(y)$ are orthogonal with respect to the measure $(1-y)^a(1+y)^b$

\be 
\int_{-1}^1 (1-y)^a(1+y)^b P_n^{(a,b)}(y) P_m^{(a,b)}(y) dy = {2^{a+b+1} \over 2n+a+b+1} {\Gamma(n+a+1)\Gamma(n+b+1) \over \Gamma(n+a+b+1) n!} \delta_{n,m}
\ee
where $\Gamma(x)$ is the Euler gamma function and $a,b>-1$. Particularly for $a=0$ we have
\be 
\int_{-1}^1 (1+y)^b P_n^{(0,b)}(y) P_m^{(0,b)}(y) dy = {2^{b+1} \over 2n+b+1} \delta_{n,m}
\ee

Also the Jacobi polynomials have the following two properties which are useful for us

\begin{align}
P_n^{(a,b)}(1) &= C_{n+a}^n \\
{d^k \over dy^k} P_n^{(a,b)}(y) &= {\Gamma(a+b+n+1+k) \over 2^k \Gamma(a+b+n+1)} P_{n-k}^{(a+k,b+k)}(y)
\end{align}

Using these facts it is easy to find that

\begin{align}
{d \over dy}\Big|_{y=1} P_n^{(0,b)}(y) &= {n(n+b+1) \over 2}\\
{d^2 \over dy^2}\Big|_{y=1} P_n^{(0,b)}(y) &= {n(n-1)(n+b+1)(n+b+2) \over 8}
\end{align}


\newpage

\bibliography{Mylib}{}

\providecommand{\href}[2]{#2}\begingroup\raggedright\begin{thebibliography}{10}

\bibitem{Moore:1991ir}
G.~W. Moore, N.~Seiberg, and M.~Staudacher, {\it {From loops to states in 2-D
  quantum gravity}},  {\em Nucl.Phys.} {\bf B362} (1991) 665--709.

\bibitem{Belavin:2008kv}
A.~Belavin and A.~Zamolodchikov, {\it {On Correlation Numbers in 2D Minimal
  Gravity and Matrix Models}},  {\em J.Phys.} {\bf A42} (2009) 304004,
  [\href{http://xxx.lanl.gov/abs/0811.0450}{{\tt arXiv:0811.0450}}].

\bibitem{Goulian:1990qr}
M.~Goulian and M.~Li, {\it {Correlation functions in Liouville theory}},  {\em
  Phys.Rev.Lett.} {\bf 66} (1991) 2051--2055.

\bibitem{Zamolodchikov:2005sj}
{Al. Zamolodchikov}, {\it {Three-point function in the minimal Liouville
  gravity}},  {\em Theor.Math.Phys.} {\bf 142} (2005) 183--196.

\bibitem{Belavin:2006ex}
{A. Belavin, Al. Zamolodchikov}, {\it {Integrals over moduli spaces, ground
  ring, and four-point function in minimal Liouville gravity}},  {\em
  Theor.Math.Phys.} {\bf 147} (2006) 729--754.

\bibitem{Polyakov:1981rd}
A.~M. Polyakov, {\it {Quantum Geometry of Bosonic Strings}},  {\em Phys.Lett.}
  {\bf B103} (1981) 207--210.

\bibitem{Knizhnik:1988ak}
V.~Knizhnik, A.~M. Polyakov, and A.~Zamolodchikov, {\it {Fractal Structure of
  2D Quantum Gravity}},  {\em Mod.Phys.Lett.} {\bf A3} (1988) 819.

\bibitem{Belavin:1984vu}
A.~Belavin, A.~M. Polyakov, and A.~Zamolodchikov, {\it {Infinite Conformal
  Symmetry in Two-Dimensional Quantum Field Theory}},  {\em Nucl.Phys.} {\bf
  B241} (1984) 333--380.

\bibitem{Zamolodchikov:2003yb}
{Al. Zamolodchikov}, {\it {Higher equations of motion in Liouville field
  theory}},  {\em Int.J.Mod.Phys.} {\bf A19S2} (2004) 510--523,
  [\href{http://xxx.lanl.gov/abs/hep-th/0312279}{{\tt hep-th/0312279}}].

\bibitem{Kazakov:1985ea}
V.~Kazakov, A.~A. Migdal, and I.~Kostov, {\it {Critical Properties of Randomly
  Triangulated Planar Random Surfaces}},  {\em Phys.Lett.} {\bf B157} (1985)
  295--300.

\bibitem{Kazakov:1986hu}
V.~Kazakov, {\it {Ising model on a dynamical planar random lattice: Exact
  solution}},  {\em Phys.Lett.} {\bf A119} (1986) 140--144.

\bibitem{Kazakov:1989bc}
V.~Kazakov, {\it {The Appearance of Matter Fields from Quantum Fluctuations of
  2D Gravity}},  {\em Mod.Phys.Lett.} {\bf A4} (1989) 2125.

\bibitem{Staudacher:1989fy}
M.~Staudacher, {\it {The Yang-Lee edge singularity on a dynamical planar random
  surface}},  {\em Nucl.Phys.} {\bf B336} (1990) 349.

\bibitem{Brezin:1990rb}
E.~Brezin and V.~Kazakov, {\it {Exactly solvable field theories of closed
  strings}},  {\em Phys.Lett.} {\bf B236} (1990) 144--150.

\bibitem{Douglas:1989ve}
M.~R. Douglas and S.~H. Shenker, {\it {Strings in Less Than One-Dimension}},
  {\em Nucl.Phys.} {\bf B335} (1990) 635.

\bibitem{Gross:1989vs}
D.~J. Gross and A.~A. Migdal, {\it {Nonperturbative Two-Dimensional Quantum
  Gravity}},  {\em Phys.Rev.Lett.} {\bf 64} (1990) 127.

\bibitem{Ginsparg:1993is}
P.~H. Ginsparg and G.~W. Moore, {\it {Lectures on 2-D gravity and 2-D string
  theory}},  \href{http://xxx.lanl.gov/abs/hep-th/9304011}{{\tt
  hep-th/9304011}}.

\bibitem{DiFrancesco:1993nw}
P.~Di~Francesco, P.~H. Ginsparg, and J.~Zinn-Justin, {\it {2-D Gravity and
  random matrices}},  {\em Phys.Rept.} {\bf 254} (1995) 1--133,
  [\href{http://xxx.lanl.gov/abs/hep-th/9306153}{{\tt hep-th/9306153}}].

\bibitem{DiFrancesco1992}
P.~Di~Francesco and D.~Kutasov, {\it {World sheet and space-time physics in
  two-dimensional (Super)string theory}},  {\em Nucl.Phys.} {\bf B375} (1992)
  119--172, [\href{http://xxx.lanl.gov/abs/hep-th/9109005}{{\tt
  hep-th/9109005}}].

\bibitem{Douglas:1989dd}
M.~R. Douglas, {\it {Strings in less than one-dimension and the generalized KdV
  hierarchies}},  {\em Phys.Lett.} {\bf B238} (1990) 176.

\bibitem{Krichever:1992sw}
I.~Krichever, {\it {The Dispersionless Lax equations and topological minimal
  models}},  {\em Commun.Math.Phys.} {\bf 143} (1992) 415--429.

\bibitem{Dubrovin:1992dz}
B.~Dubrovin, {\it {Integrable systems in topological field theory}},  {\em
  Nucl.Phys.} {\bf B379} (1992) 627--689.

\bibitem{Dubrovin1996} B.~Dubrovin, Geometry of 2D topological field theories,
in: Integrable Systems and Quantum Groups, Montecatini, Terme, 1993.
Editors: M.Francaviglia, S. Greco. Springer Lecture Notes in Math.
{\bf 1620} (1996), 120--348.

\bibitem{Segal1985}
G.~Segal and G.~Wilson, {\it {Loop groups and equations of KdV type}},  {\em
  Pub. Math IHES} {\bf 61} (1985) 5.

\bibitem{Dijkgraaf:1990dj}
R.~Dijkgraaf, H.~L. Verlinde, and E.~P. Verlinde, {\it {Topological strings in
  d less than 1}},  {\em Nucl.Phys.} {\bf B352} (1991) 59--86.

\bibitem{Ginsparg:1990zc}
P.~H. Ginsparg, M.~Goulian, M.~Plesser, and J.~Zinn-Justin, {\it {(p, q) String
  actions}},  {\em Nucl.Phys.} {\bf B342} (1990) 539--563.


\bibitem{GD76} I.M.~Gelfand, L.A.~Dickey, {\it Fractional powers of operators and Hamiltonian systems} {\em Funct. Anal. Appl.} {\bf 10:4} (1976) 259-273.

\bibitem{adler} M.~Adler, {\it On a trace functional for formal pseudodifferential operators and the symplectic structure of Korteweg--de Vries equations}, {\em Inventiones Math.} {\bf 50} (1979) 219-248.


\end{thebibliography}\endgroup

\end{document}